\documentclass[english,prl,twocolumn,aps,amssymb,nofootinbib]{revtex4-1}
\usepackage[T1]{fontenc}
\usepackage{amsmath}
\usepackage{amssymb}
\usepackage{graphicx}
\usepackage{braket}
\usepackage{babel}
\usepackage{color}
\usepackage{braket}
\usepackage{xcolor,soul}

\usepackage[T1]{fontenc}
\usepackage[utf8]{inputenc}
\usepackage{textcomp}
\usepackage{color}

\usepackage{babel}
\usepackage[symbol]{footmisc}

\makeatother

\begin{document}

\title{Month-long-lifetime microwave spectral holes in an erbium-doped scheelite crystal at millikelvin temperature}

%\title{Week-long-lived nuclear-spin frozen core probed by microwave spectral hole burning in a paramagnetic-rare-earth-ions-doped crystal}

\author{Z. Wang$^{  1,2}$, S. Lin$^{  3}$, M. Le Dantec$^{  1}$, M. Rancic$^{  1}$, P. Goldner$^{4}$, S. Bertaina$^{5}$, T. Chaneliere$^{6}$, R. B. Liu$^{  3}$, D. Esteve$^{1}$, D. Vion$^{1}$, E. Flurin$^{1}$, P. Bertet$^{1}$}

\affiliation{
\\ \
\\
$^1$Quantronics group, Universit\'e Paris-Saclay, CEA, CNRS, SPEC, 91191 Gif-sur-Yvette Cedex, France\\
$^2$D\'epartement de Physique et Institut Quantique, Universit\'e de Sherbrooke, Sherbrooke, Qu\'ebec, Canada\\
$^3$Department of Physics, Centre for Quantum Coherence,\\
The Hong Kong Institute of Quantum Information Science and Technology,\\
and New Cornerstone Science Laboratory, The Chinese University of Hong Kong,
Shatin, New Territories, Hong Kong, China\\
$^4$Chimie ParisTech, PSL University, CNRS, Institut de Recherche de Chimie Paris, 75005 Paris, France\\
$^5$CNRS,  Aix-Marseille  Universit\'e,  IM2NP  (UMR  7334),  Institut  Mat\'eriaux Micro\'electronique  et  Nanosciences de Provence,  Marseille, France \\
$^6$Univ. Grenoble Alpes, CNRS, Grenoble INP, Institut N\'eel, 38000 Grenoble, France}

\date{\today}

\maketitle

%Rare-earth--ions (REI) doped crystals have remarkable optical properties, characterized by narrow homogeneous linewidths, combined with a sizeable inhomogeneous broadening. Spectral hole burning has been widely applied for high-resolution spectroscopy 
%High-power optical irradiation frequently leads to the pumping of the resonant ions into other ground-state sub-levels causing the appearance of a hole in the optical ensemble linewidth, a phenomenon known as spectral hole-burning which is widely used for high-resolution optical spectroscopy. 
%
\textbf{
%Spectral hole burning is a cornerstone of high-resolution spectroscopy, particularly useful for the study of rare-earth-ions-doped crystals optical properties. Here, we report the observation of an unusual hole-burning phenomenon, at microwave frequency and millikelvin temperatures. 
%Rare-earth--ions (REI) doped crystals have remarkable optical and spin properties characterized by narrow homogeneous linewidths, which can be accessed despite the large inhomogeneous broadening of the ensemble line through spectral hole burning. Here, we report hole burning spectroscopic measurements of the spin transition of a paramagnetic REI (Er$^{3+}$) in a scheelite crystal of CaWO4, at microwave frequency and millikelvin temperatures. The hole-burning mechanism is the progressive polarization of the $^{183}$W nuclear spins that surround the ions, caused by the microwave pump tone. Narrow hole widths of $\sim 2$\,kHz are measured, revealing long homogeneous electron spin coherence times. The repeated application of pairs of microwave pulses generates a periodic modulation of the Er$^{3+}$ density profile, which we observe spectrally and in the time-domain as an accumulated echo. The lifetime of the holes as well as of the accumulated echoes rises steeply as the sample temperature is decreased. It reaches one week at the lowest temperature of 10mK, indicating nearly complete suppression of long-term electron spin spectral diffusion and quasi-infinite lifetime of the $^{183}$W nuclear spins in the Er$^{3+}$ vicinity. 
Rare-earth--ion (REI) ensembles in crystals have remarkable optical and spin properties characterized by narrow homogeneous linewidths relative to the inhomogeneous ensemble broadening. This makes it possible to precisely tailor the ensemble spectral density and therefore the absorption profile by applying narrow-linewidth radiation to transfer population into auxiliary levels, a process broadly known as spectral hole burning (SHB)~\cite{macfarlane_optical_2007}. REI-doped crystals find applications in information processing, both classical (pattern recognition~\cite{cole_coherent_2002}, filtering~\cite{li_pulsed_2008}, spectral analysis~\cite{berger_rf_2016}) and quantum (photon storage~\cite{afzelius_multimode_2009,clausen_quantum_2011,putz_spectral_2017,lago-rivera_telecom-heralded_2021}), all protocols requiring suitable ensemble preparation by SHB as a first step. In $\mathrm{Er}^{3+}$-doped materials, the longest reported hole lifetime is one minute~\cite{rancic_coherence_2018}, and longer lifetimes are desirable.  
%Erbium-doped crystals are particularly interesting for their optical transition in the telecom window; recent measurements reported minutes-long-lived holes by optical SHB and storage in the hyperfine levels of 167Er in YSO. 
Here, we report SHB and accumulated echo~\cite{hesselink_picosecond_1979,hesselink_photon_1981} measurements in a scheelite crystal of CaWO4 by pumping the electron spin transition of $\mathrm{Er}^{3+}$ ions at microwave frequencies and millikelvin temperatures, with nuclear spin states of neighboring $^{183}\mathrm{W}$ atoms serving as the auxiliary levels. 
%Narrow hole widths of $\sim 2$\,kHz are measured, revealing long homogeneous electron spin coherence times. 
The lifetime of the holes and accumulated echoes rises steeply as the sample temperature is decreased, exceeding a month at 10mK.
%. It reaches one week at the lowest temperature of 10mK, indicating nearly complete suppression of long-term electron spin spectral diffusion and quasi-infinite lifetime of the $^{183}$W nuclear spins in the Er$^{3+}$ frozen core []. 
Our results demonstrate that millikelvin temperatures can be beneficial for signal processing applications requiring long spectral hole lifetimes.}

Spectral holes were first observed by Feher by selectively saturating a portion of the inhomogeneously broadened spin resonance line of donors in silicon with a microwave pump tone~\cite{feher_electron_1959}. SHB has since then been an important tool in Electron Paramagnetic Resonance (EPR) spectroscopy, forming for instance the basis for an all-microwave hyperfine spectroscopy sequence ~\cite{wacker_fourier_1991,schosseler_pulsed_1994}. In the optical domain, SHB has been a cornerstone of REI-doped crystals spectroscopy, particularly helpful for resolving the hyperfine structure below the inhomogeneous linewidth ~\cite{erickson_optical_1977,shelby_measurement_1978,macfarlane_optical_2007}. 

Besides spectroscopy, SHB is also useful to prepare a desired absorption profile into the REI ensemble line, by transfering part of the population into auxiliary storage levels in a frequency-selective manner. The storage levels are usually the REI nuclear spin hyperfine levels~\cite{shelby_optical_1980}, or more rarely the spin states of neighboring nuclei of the host crystal~\cite{macfarlane_optical_1981}. The spectral preparation constitutes the first step of many information processing protocols that rely on REI ensembles (for instance, atomic frequency combs for optical quantum memories ~\cite{afzelius_multimode_2009}). In these applications, the lifetime of the hole/anti-hole pattern can be a limiting factor and has therefore been a focus of recent studies~\cite{cruzeiro_efficient_2018}. Lifetimes of days or weeks have been reported by optically pumping hyperfine levels of non-paramagnetic REI systems such as $\mathrm{Eu}^{3+}\mathrm{:YSO}$~\cite{konz_temperature_2003}. Shorter lifetimes have also been observed in paramagnetic Kramers REIs; with lifetimes ranging from a few seconds to a minute reported for optically pumped electronic and hyperfine transitions of Nd$^{3+}$ and $^{167}\mathrm{Er}^{3+}$ doped YSO, respectively~\cite{cruzeiro_spectral_2017,rancic_coherence_2018}.
%, of order a few seconds by pumping optically and storing on the paramagnetic transition of $\mathrm{Nd}^{3+}\mathrm{:YSO}$~\cite{cruzeiro_spectral_2017}, and of order a minute by pumping optically and storing in the hyperfine levels of the $^{167}\mathrm{Er}^{3+}\mathrm{:YSO}$ ground state~\cite{rancic_coherence_2018}. 
We note that $\mathrm{Er}^{3+}$ is a particularly interesting REI for applications, owing to its $1.5\mu\mathrm{m}$ optical transition in the c-band telecom window. Here, we report hole lifetimes as long as one month in $\mathrm{Er}^{3+}\mathrm{:CaWO}_4$ at $10$\,mK by pumping at microwave frequency on the paramagnetic transition, and using the spin states of neighboring $^{183}$W nuclei of the host crystal as auxiliary levels.

\begin{figure}
\centerline{\includegraphics[width=0.5\textwidth]{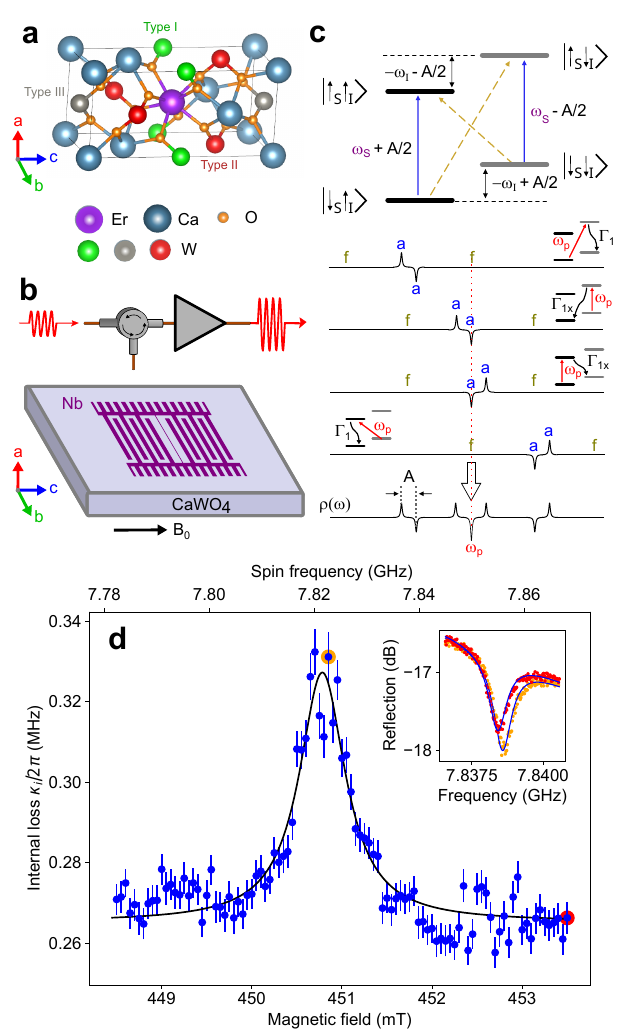}}
\caption{
\textbf{Principle of the experiment}. \textbf{a}. Schematic of the scheelite unit cell, showing an Er$^{3+}$ ion (purple) substituting for a Ca$^{2+}$ ion (blue). When the magnetic field $B_0$ is applied along the $c$ axis, nearest neighbor $^{183}$W  nuclei belong to $2$ sets of four sites with identical hyperfine coupling, which we call Type I (in green), and Type II (in red). \textbf{b}. Schematic of the setup. A niobium lumped-element resonator is deposited on top of a $\mathrm{CaWO_4}$ sample, and probed in microwave reflectometry. \textbf{c}. Energy levels of an Er$^{3+}$ electron spin of frequency $\omega_S$ with levels $\downarrow_S$ (ground state) and $\uparrow_S$ (excited state), coupled to a nuclear spin of frequency $\omega_I$ and levels $\uparrow_I$ (ground state) and $\downarrow_I$ (excited state). EPR-allowed transitions (noted $a$) are at frequency $\omega_S \pm A/2$, and EPR-forbidden transitions (noted $f$) at $\omega_S \pm \omega_I$. Irradiation at $\omega_p$ can be resonant with all $4$ possible transitions of spin packets with different spin frequency $\omega_S$ due to inhomogeneous broadening, resulting in a characteristic Hole/Anti-hole pattern in the spin density $\rho(\omega)$. \textbf{d}. Blue dots : measured resonator internal losses $\kappa_i$ as a function of magnetic field $B_0$, showing a peak when the Er$^{3+}$ ions are resonant with $\omega_0$. Solid black line is a Lorentzian fit, yielding the inhomogeneous linewidth $0.62$\,mT translating in a frequency $\Gamma/2\pi = 10.8$\,MHz. Inset: reflection coefficient $S_{11}(\omega)$ away from (red dots) and at (orange) the Er$^{3+}$ resonance, with Lorentzian fits. }
\label{fig1}
\end{figure}

The unit cell of scheelite is depicted in Fig.~\ref{fig1}a. It has a tetragonal symmetry, with two axes ($a,b$) equivalent under a $90^\circ$ rotation around the $c$ axis. $\mathrm{Er}^{3+}$ ions substitute for $\mathrm{Ca}^{2+}$, as is often the case for REIs in scheelite. $\mathrm{Er}^{3+}$ is a Kramers doublet, and at low temperatures only the lowest-energy doublet is populated, so that $\mathrm{Er}^{3+}$ ions behave as an effective electron spin $S=1/2$, with energy levels denoted as $\downarrow_S$ (ground state) and $\uparrow_S$ (excited state). Note that we will only consider the $\mathrm{Er}^{3+}$ isotopes which have a zero nuclear spin. Owing to the $S_4$ symmetry of this site, $\mathrm{Er}^{3+}:\mathrm{CaWO}_4$ has a gyromagnetic tensor $\gamma$ that is diagonal along the crystalline $a,b,c$ axes, with $\gamma_{||} / 2\pi = 17.35$\,GHz/T, and  $\gamma_{\perp} / 2\pi = 117$\,GHz/T, depending on whether the magnetic field $B_0$ is applied parallel or perpendicular to the $c$ axis~\cite{antipin_a._paramagnetic_1968}. Most atoms in the scheelite lattice have zero nuclear spin, apart from tungsten, whose $14\%$-abundant isotope $^{183}\mathrm{W}$ has a nuclear spin $I=1/2$ with a gyromagnetic ratio $\gamma_W/2\pi = 1.77394$\,MHz/T~\cite{knight_solid-state_1986}. 

The measurements are conducted at $10$\,mK, by microwave reflectometry on a superconducting resonator (resonance frequency $\omega_0/2\pi = 7.839$\,GHz) patterned directly on top of a scheelite crystal (see Fig.~\ref{fig1}b), with an erbium concentration of $3$\,ppb measured by CW EPR~\cite{le_dantec_electron_2022}. The resonator consists of a finger capacitor in parallel with a $\mu \mathrm{m}$-wide wire, acting as the inductance, and oriented approximately along the crystalline $c$ axis. A magnetic field $B_0$ is applied in the plane of the resonator, approximately along the crystalline $c$ axis. Measuring the reflection coefficient $S_{11}(\omega)$ yields the resonator energy coupling rate $\kappa_c \sim 3\cdot 10^7 \,\mathrm{s}^{-1}$ and internal loss rate $\kappa_i$. The presence of the $\mathrm{Er}^{3+}$ ions manifests itself as an increase in $\kappa_i(B_0)$ when the spin resonance frequency $\omega_S = |S \cdot \gamma \cdot B_0|$ is resonant with $\omega_0$ (see Fig.~\ref{fig1}); the measured peak width of $ 0.62$\,mT translates into an inhomogeneous linewidth $\Gamma/2\pi = 10.8$\,MHz. The inhomogeneous broadening is due to the electrostatic and magnetic local environment of each $\mathrm{Er}^{3+}$ ion which varies throughout the crystal~\cite{mims_broadening_1966,le_dantec_twenty-threemillisecond_2021,billaud_electron_2024}. In the following, $B_0$ is fixed at the center of the spin resonance, with the coils in persistent mode for increased stability (see Methods). The spin density $\rho(\omega)$ is obtained by measuring the reflection coefficient $S_{11}(\omega)$ with a vector network analyzer (see Methods). Note that since $\Gamma \gg \kappa_c$, the equilibrium density $\rho_0$ can and shall be considered constant across the resonator linewidth. Moreover, most experiments reported here are conducted at temperatures $T$ satisfying $k T \ll \hbar \omega_S$ such that at thermal equilibrium, only $\ket{\downarrow_S}$ is populated significantly. 

SHB in our sample arises because of the magnetic dipolar interaction between each $\mathrm{Er}^{3+}$ ion and the $^{183}\mathrm{W}$ nuclear spins surrounding it. Consider for simplicity the coupling of one $\mathrm{Er}^{3+}$ ion spin (operator $S$) to one $^{183}\mathrm{W}$ nuclear spin (operator $I$), described in the secular approximation by the Hamiltonian $H = \omega_S S_z + \omega_I I_z + S_z (A I_z + B I_x)$. Here, $\omega_I = - \gamma_W B_0$ is the $^{183}\mathrm{W}$ Larmor frequency, and $A$ (resp. $B$) is the isotropic (resp., anisotropic) component of the hyperfine interaction. In the limit where $|\omega_I| \gg A,B$, the energy eigenstates are close to the uncoupled states $\ket{\downarrow_S \uparrow_I},\ket{\downarrow_S \downarrow_I},\ket{\uparrow_S \uparrow_I},\ket{\uparrow_S \downarrow_I}$ (in increasing energy order). The nuclear-spin-conserving transitions $\ket{\downarrow_S \uparrow_I} \leftrightarrow \ket{\uparrow_S \uparrow_I}$ and $\ket{\downarrow_S \downarrow_I} \leftrightarrow \ket{\uparrow_S \downarrow_I}$ at respectively $\omega_S - A/2$ and $\omega_S + A/2$ are EPR-allowed with $\langle \downarrow_S \uparrow_I | S_x | \uparrow_S \uparrow_I \rangle \approx \langle \downarrow_S \downarrow_I | S_x | \uparrow_S \downarrow_I \rangle \approx 1/2 $. Due to a slight electron-spin-state-dependent mixing of the nuclear spin states caused by the $B S_z I_x$ term, the forbidden transitions $\ket{\downarrow_S \uparrow_I} \leftrightarrow \ket{\uparrow_S \downarrow_I}$ and $\ket{\downarrow_S \downarrow_I} \leftrightarrow \ket{\uparrow_S \uparrow_I}$, at respective frequencies $\omega_H = \omega_S - \omega_I$ and $\omega_L = \omega_S + \omega_I$, can also be weakly driven since $|\langle \downarrow_S \uparrow_I | S_x | \uparrow_S \downarrow_I \rangle | \approx | \langle \downarrow_S \downarrow_I | S_x | \uparrow_S \uparrow_I \rangle | \approx |B/(4 \omega_I)| $.
%, albeit with a much lower $S_x$ matrix element, approximately equal to $B/(4 \omega_I)$. 
SHB relies on relaxation pathways of the electron-nuclear spin system. At low temperature, electron-spin relaxation occurs via spontaneous emission of phonons through the direct process, and of microwave photons through the Purcell effect~\cite{bienfait_controlling_2016}. The measurements reported here were performed at a microwave power (-106 dBm at sample input) large enough to excite a large number of $\mathrm{Er}^{3+}$ ion spins in the bulk of the crystal. The average radiative rate of these ions, located at $\sim 100 \mu \mathrm{m}$ from the inductance, is negligible compared to their non-radiative rate dominated by the direct phonon emission process~\cite{le_dantec_twenty-threemillisecond_2021}, in contrast with recent detection of single-$\mathrm{Er}^{3+}$ spins which were very close ($\sim 100$\,nm) to the inductance~\cite{wang_single_2023}. Therefore, one can assume an identical non-radiative relaxation rate for all the ions, measured to be $\Gamma_{1} \approx 5\,\mathrm{s}^{-1}$ in the conditions of our experiment~\cite{le_dantec_electron_2022}. Because of the nuclear-electron-spin mixing, electron spin relaxation with nuclear-spin-flipping is moreover also possible via the forbidden transitions, at a much lower rate $ \Gamma_{1x} \approx \Gamma_{1} B^2/(4 \omega_I^2)$.

SHB occurs by applying a strong pump tone at frequency $\omega_p$, of duration much longer than the characteristic electronic relaxation times $\Gamma_{1}^{-1}$ and $\Gamma_{1x}^{-1}$. Due to inhomogeneous broadening, the pump is resonant with each of the $4$ transitions for $4$ different spin packets (see Fig.~\ref{fig1}). When the pump is resonant with a forbidden transition (say, $\ket{\downarrow_S \uparrow_I} \leftrightarrow \ket{\uparrow_S \downarrow_I}$, occurring when $\omega_p = \omega_H$), it drives this transition, which is rapidly followed by electron spin relaxation into the $\downarrow_S \downarrow_I$ state. This transfers spin population of this packet from $\omega_p - (\omega_I + A/2)$ into $\omega_p - (\omega_I - A/2)$ (see Fig.~\ref{fig1}c). When the pump is resonant with an allowed transition (say, $\ket{\downarrow_S \uparrow_I} \leftrightarrow \ket{\uparrow_S \uparrow_I}$, when $\omega_p = \omega_S - A/2$), the system will cycle many times on the allowed transition, until one cross-relaxation event happens into the state $\ket{\downarrow_S \downarrow_I}$, which transfers population from $\omega_p$ into $\omega_p + A$. Summing up all $4$ spin packets yields a predicted H/AH pattern shown in Fig.~\ref{fig1}c. Once a H/AH pattern has been burned by a strong pump at $\omega_p$, it can be probed with a much weaker ($\sim 60$\,dB less) probe tone of varying frequency; it can also be washed out by sweeping the frequency of a strong pump tone, enabling to effectively reset the spin density.

\begin{figure}
\centerline{\includegraphics[width=0.5\textwidth]{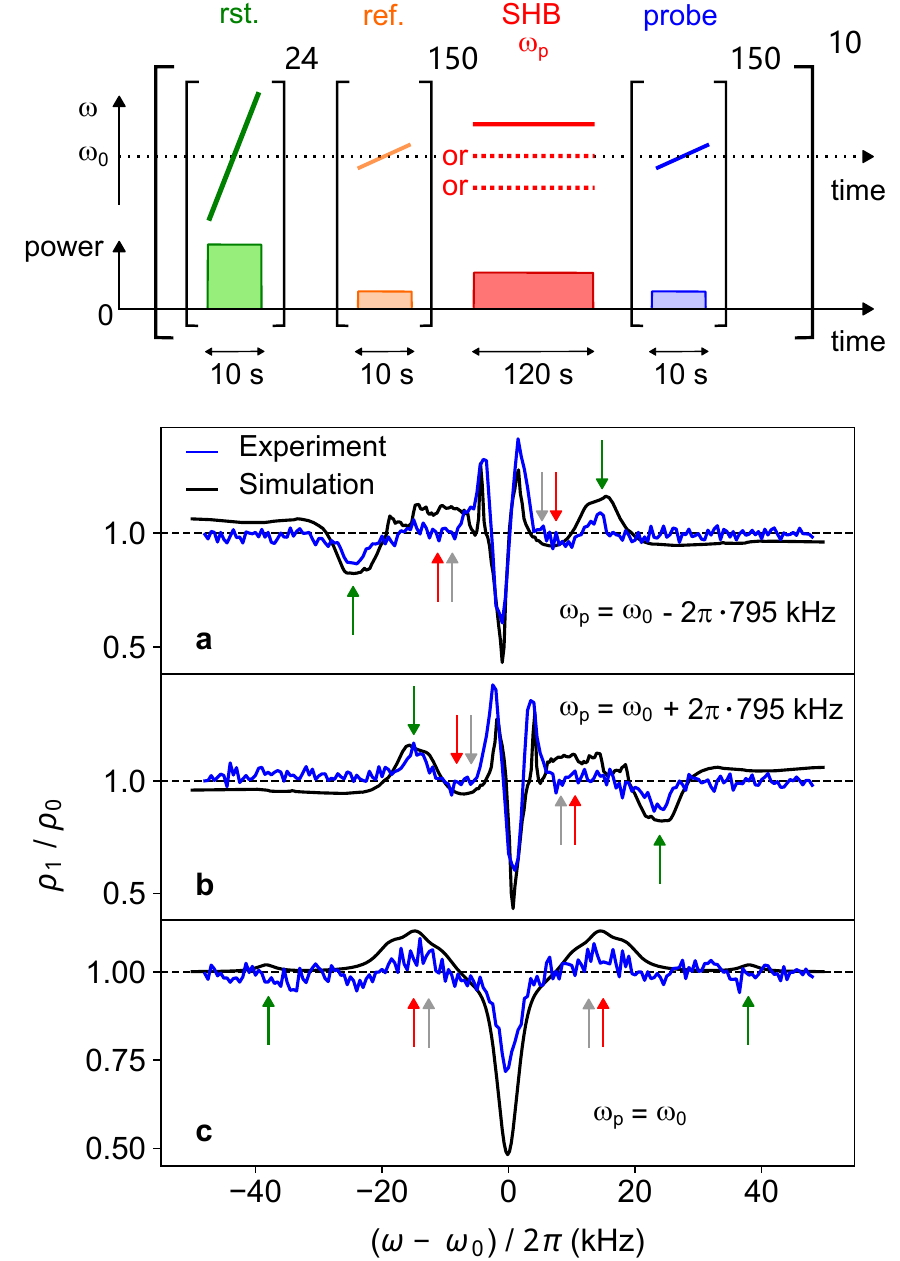}}
\caption{
\textbf{Spectral hole burning.} Measured (blue solid line) and simulated (black solid line) spin density $\rho_1(\omega)$ normalized to the unperturbed spin density $\rho_0(\omega)$. Each sequence starts with a repetitive high power ($\sim -96$\,dBm) reset scan with a range of 4 MHz centered at $\omega_0$, followed by a lower power ($\sim -156$ \,dBm) reference scan yielding $\rho_0(\omega)$ , a pumping step at single frequency $\omega_p$ (power -106 dBm), and a measurement scan yielding $\rho_1(\omega)$. The entire sequence is repeated for 10 times. The pump frequency $\omega_p/2\pi$ is $\omega_0 /2\pi - 795$\,kHz (a), $\omega_0 /2\pi + 795$\,kHz (b), and $\omega_0$ (c). Green, red and grey arrows point respectively to H/AH structure due to Type I, Type II and Type III $^{183}\mathrm{W}$ nuclear spins.}
\label{fig2}
\end{figure}

We record SHB spectra with the following pulse sequence. We first use high-power scans to reset the spin density to its equilibrium value, $\rho_0(\omega)$, probed by weak scans as a reference. We then apply a $120$\,s-long microwave pulse at frequency $\omega_p$ with intermediate power for burning spectral holes. We finally measure the resulting spin density $\rho_1(\omega)$ with low-power scans. The experiments are repeated 10 times and averaged. One difficulty is that the complete SHB pattern does not fit inside the narrow linewidth of our resonator, where $\rho(\omega)$ can be measured. So, we chose to record and erase subsequently 3 spectra around the line frequency center $\omega_0$ after applying a pump at respectively $\omega_0 + \omega_I$ (red forbidden transition pumping), $\omega_0 - \omega_I$ (blue forbidden transition pumping), and $\omega_0$ (allowed transition pumping).

%In this set of measurements, the magnetic field $B_0$ was as accurately aligned as possible with the crystalline c-axis (see Methods).

The relative density change, $\rho_1(\omega)/\rho_0(\omega)$, is plotted in Fig.~\ref{fig2} for the $3$ pump frequencies. Consider first the forbidden transition spectrum. We focus on the red-sideband (Fig.~\ref{fig2}a), since the blue-sideband spectrum is a mirror image as expected. We first note that the frequency difference between the pump and the center of the hole of $795$\,kHz is close to the unperturbed $^{183}\mathrm{W}$ Larmor frequency, $|\omega_I| /2\pi = 794$kHz, confirming that the SHB occurs by storage in $^{183}\mathrm{W}$ nuclear spin states. Two sets of H/AH patterns can be identified in the spectrum, marked by green (resp. red) arrows. From comparison to computed dipolar magnetic couplings, we identify the green set with the largest isotropic hyperfine coupling $|A_I|/2\pi \sim 35$\,kHz as arising from the four $^{183}\mathrm{W}$ nuclear spins located on the (ab) plane-parallel square surrounding the $\mathrm{Er}^{3+}$ site (Type I in Fig.~\ref{fig1}a). The red set of shallower peaks/dips with $|A_{II}|/2\pi \sim 20$\,kHz is attributed to the four type II $^{183}\mathrm{W}$ nuclear spins (see Fig.~\ref{fig1}a). At the center of the spectrum, a narrow $2$-kHz-wide hole flanked with two narrow side anti-holes is observed, which we attribute to the numerous weakly-coupled nuclear spins located outside of the unit cell. The low $2$-kHz linewidth indicates low spectral diffusion over the time scales of the pumping and of the probing; it is in agreement with long coherence times measurements in $\mathrm{Er}^{3+}\mathrm{:CaWO}_4$ ~\cite{le_dantec_twenty-threemillisecond_2021,wang_single_2023}. A simulation reproduces the observations satisfyingly (see Fig.~\ref{fig2} and Methods), the smaller measured modulation amplitude may be linked to magnetic field drifts during the pumping sequence. Note that the agreement was obtained by adjusting the values of Type I and Type II spins hyperfine constants to the data; the fitted values differ from the pure dipolar contribution, likely pointing to the existence of a non-zero Fermi contact interaction, as already observed by Mims in $\mathrm{Ce}^{3+}\mathrm{:CaWO}_4$~\cite{mims_pulsed_1997}.

The resonant spectrum shows a central hole, flanked with two broad and shallow anti-holes peaking at $\sim 17$\,kHz. Those are attributed to the Type II $^{183}\mathrm{W}$ spin. The two Type I $^{183}\mathrm{W}$ anti-holes expected at $\pm 35$\,kHz are not visible, due to their low $B$ value when $B_0$ is well aligned with the $c$ axis. This is confirmed by the simulations, which indeed predict un-measurably small values for the Type I anti-holes in the resonant spectrum (see Fig.~\ref{fig2}).

%One may wonder why two anti-hole at $\pm 40$\,kHz are not observed from the Type I spins, as would be expected from the qualitative discussion earlier. We believe this is due to the low value of the hyperfine coupling $B$ for the Type I spins when $B_0$ is well aligned with the $c$ axis. As a result, cross-relaxation of other nuclear spins occurs faster than for the Type I spins, leading to the absence of the anti-hole feature in the resonant spectrum. This is confirmed by the simulations, which indeed predict an un-measurably small value for the Type I anti-holes (see Fig.~\ref{fig2}). 
\begin{figure}
\centerline{\includegraphics[width=0.5\textwidth]{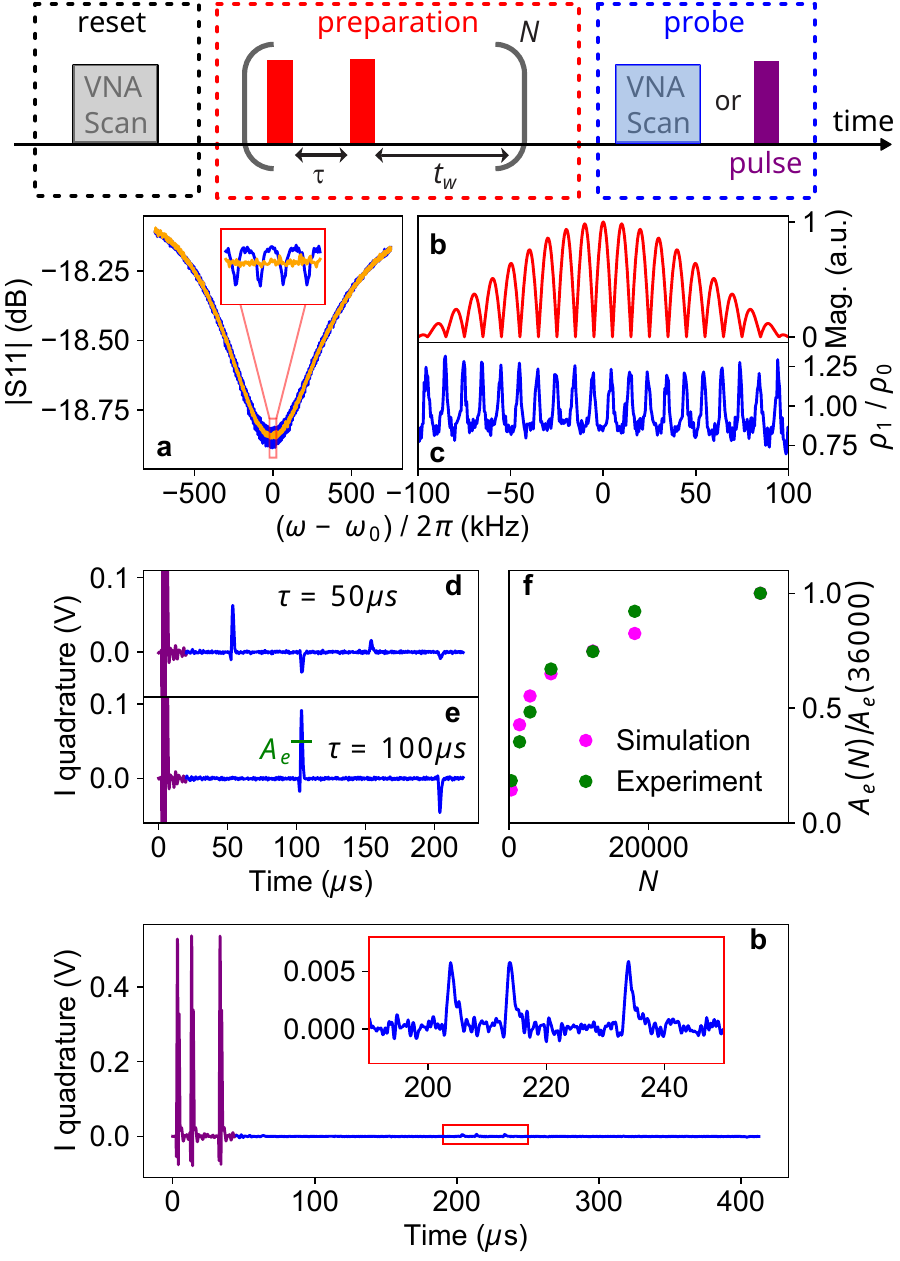}}
\caption{
\textbf{Two-pulse pumping and accumulated echo.} Pumping consists of two consecutive pulses of frequency $\omega_0$ and duration $1 \, \mu \mathrm{s}$, separated by a delay $\tau$, and repeated $N$ times after a waiting time $200$\,ms. \textbf{a}. Measured reflection coefficient without (orange line) and with (blue line) two-pulse pumping. \textbf{b}. Computed spectral power density of the two pulses; it goes to zero at frequencies $\tilde{\omega}_k$ (see text). \textbf{c}. Measured $\mathrm{Er}^{3+}$ density $\rho_1(\omega)$ after two-pulse pumping, renormalized to the reference density $\rho_0(\omega)$. It shows a periodic pattern of anti-holes at frequencies $\tilde{\omega}_k$. \textbf{d,e}. The blue line is the measured average value of the $I$ quadrature following a microwave pulse (purple part) at $\omega_p$, after a periodic grating was generated by two-pulse pumping with $\tau = 50 \mu \mathrm{s}$ (d) and $\tau = 100 \mu \mathrm{s}$ (e). \textbf{f}. First-echo amplitude $\mathrm{A_e}$ as a function of $N$. green (pink) dots are measurements (simulations). \textbf{g}. The blue line is the measured average value of the $I$ quadrature following three consecutive microwave pulses (purple) at $\omega_0$, after a periodic grating was generated by two-pulse pumping with $\tau = 200 \mu \mathrm{s}$. The inset is a zoom on the $3$ generated echoes.}
\label{fig3}
\end{figure}

We now study the case when the pump consists of pairs of short microwave pulses at $\omega_0$, separated by a delay $\tau$, and repeated a large number of times $N$ (see Fig.~\ref{fig3}). A qualitative understanding can be obtained by considering the power spectrum of a pair of pump pulses, which is proportional to $\cos^2 [2\pi (\omega - \omega_0) / \tau] $ and vanishes at all frequencies $\tilde{\omega}_k = \omega_0 + \pi (2k + 1) / \tau$, $k$ being an integer (see Fig.~\ref{fig3}b). Consequently, all $\mathrm{Er}^{3+}$ ions except those at these frequencies have a finite probability to be excited and finally cross-relax with a $^{183}\mathrm{W}$ nuclear spin flipping, as explained above. This leads to a progressive population build-up at $\tilde{\omega}_k$, and a corresponding depletion of all other frequencies, forming an effective spectral grating~\cite{hesselink_photon_1981}. This accumulated grating measured as a change in spin density resulting from the pulse pair pumping is shown in Fig.~\ref{fig3}c in a narrow frequency range around $\omega_0$, with the reflection coefficient $S_{11}(\omega)$ shown on a larger scale in Fig.~\ref{fig3}a. Both show the expected modulation.

The spectral grating in the spin density can also be probed in the time domain, by measuring the spin response following a short excitation probe pulse. As shown in Fig.~\ref{fig3}d-e, a series of echoes are observed, at times $j \tau$ ($j$ positive integer), with alternating phases, and an overall decaying amplitude $A_{e,j}$. Such accumulated echoes were also observed in optics~\cite{hesselink_picosecond_1979}; they form the basis for the generation of atomic frequency combs~\cite{afzelius_multimode_2009}. The echo emission is readily understood by the fact that the Free-Induction Decay emission from a spin ensemble is the Fourier Transform of the spin density, so that the periodic modulation in the frequency domain corresponds to a delay-line effect in the time domain. The emission of several echoes is related to the grating harmonics; the echo train is well-reproduced by a simple computation of the measured spin density Fast Fourier Transform (see Methods). The progressive buildup of the modulation is studied in Fig.~\ref{fig3}f, in which we plot the amplitude of the first accumulated echo $A_e$ as a function of the number of pump pairs of pulses. A plateau is reached after $N \sim  2 \cdot 10^4 $. A similar buildup dynamics is seen in the simulations, indicating that they capture the main features of the pumping mechanism. We finally show in Fig.~\ref{fig3}g that the delay-line effect works also when several input pulses are present, similar to atomic frequency comb memories~\cite{afzelius_multimode_2009}.

We now measure the lifetime of the holes imprinted in the spin density. We show here results for the accumulated echoes (similar results were obtained for the hole burning, see Methods). A pumping sequence consisting of $N=18000$ pairs of pulses separated by $\tau = 100\,\mu \mathrm{s}$ is applied to the spins, with a waiting time $t_w = 0.2$\,s. A probe pulse is then sent to the spins, and the accumulated echo amplitude is recorded as a function of the time $t$ elapsed since the end of the pumping sequence. Echoes are still measurable after $6$ days of waiting time. 

\begin{figure}[h!]
\centerline{\includegraphics[width=0.5\textwidth]{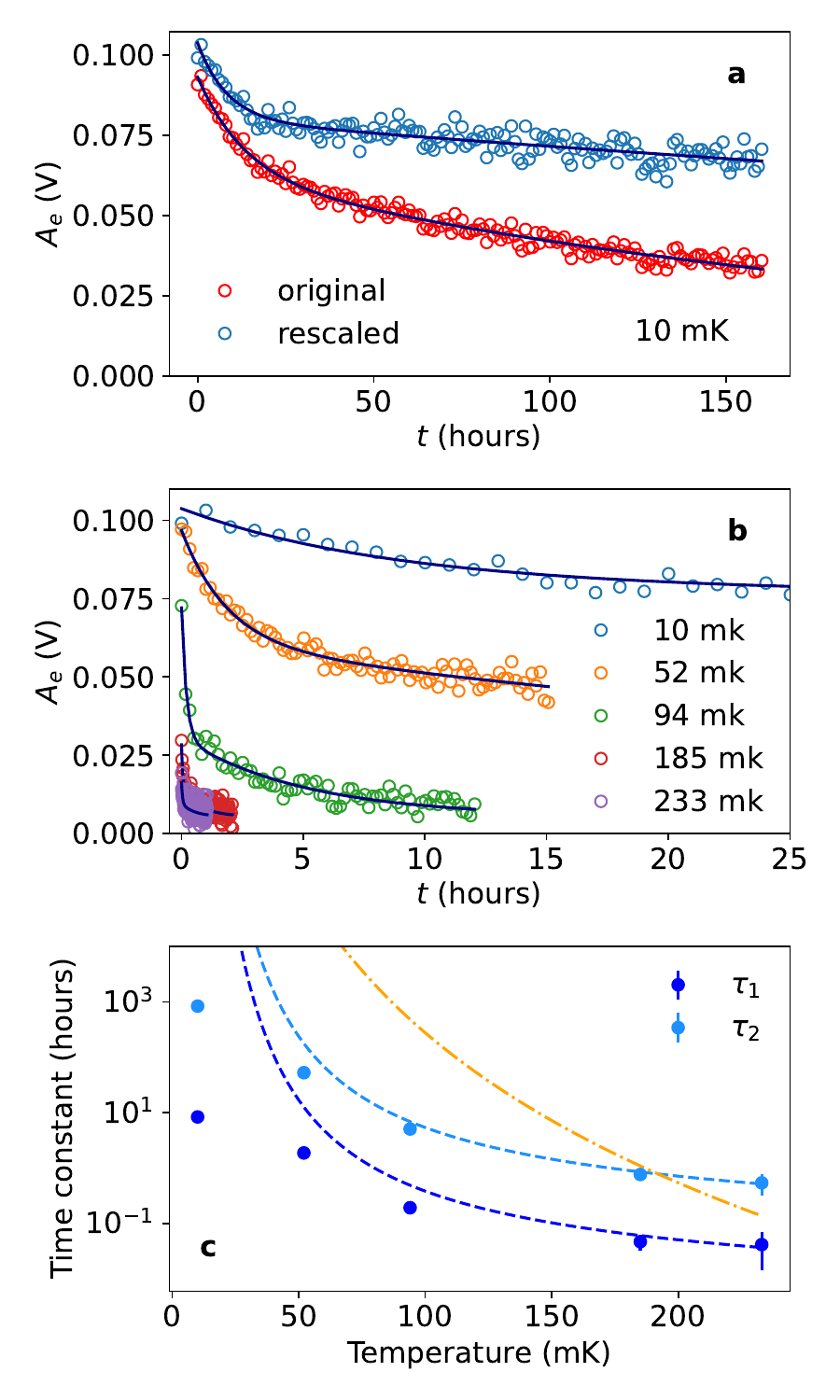}}
\caption{
\textbf{Temperature dependence of accumulated echo.} \textbf{a}. Red circles show the measured average amplitude of the first accumulated echo, after a delay $t$ following the generation of a periodic grating by two-pulse pumping with $\tau = 100 \mu \mathrm{s}$ and $N=18000$, at $10$\,mK. Blue circles show the rescaled echo amplitude. The solid line is a fit by a sum of two exponentially decaying functions, 
yielding $\tau_1 = 8.4$\,h and $\tau_2 = 841.9$\,h for rescaled data. \textbf{b}. Measured rescaled echo amplitude under various sample temperatures. \textbf{c}. Measured time constants $\tau_1$ (blue circles) and $\tau_2$ (light blue circles) as a function of the sample temperature, $T$. Fitting the data between 50\,mK and 220\,mK with the Orbach formula yields $\Gamma_{1x}=0.002$\,s$^{-1}$ ($\tau_1$, dark blue dashed line) and 0.03\,s$^{-1}$ ($\tau_2$, light blue dashed line). The orange dash-dot line shows $\sim T^{-9}$ scaling, as expected for the Raman process. 
}
\label{fig4}
\end{figure}

In these measurements, the probe pulses themselves have an impact on the spin density modulation by exciting the erbium ions and enabling cross-relaxation, artificially accelerating the decay. We calibrate this spurious effect by recording the echo amplitude as a function of the number of probe pulses in the same conditions of pumping and probing, but without any waiting time between the pulses. The echo decay at $10$\,mK, rescaled by this reference curve, is shown in Fig.~\ref{fig4}a (see Methods). The long-time decay is considerably slowed down compared to the non-rescaled data, indicating that the probe pulse impact was actually the dominant source of decay, despite the low rate of probing (five pulses every $60$\, minutes). The measurements are repeated at various sample temperatures, and the rescaled echo decay curves are shown in Fig.~\ref{fig4}b. All curves display a non-exponential decay, which we phenomenologically fit by a sum of $2$ exponentials, $A_1 \mathrm{e}^{- t / \tau_1} + A_2 \mathrm{e}^{- t / \tau_2}$, with $\tau_1$  ($\tau_2$) the short (long) time constant. The time constants $\tau_1$ and $\tau_2$ are shown as a function of the temperature $T$ in Fig.~\ref{fig4}c. A strong dependence is observed, with both time constants sharply decreasing when the temperature is increased, $\tau_2$ in particular changing by three orders of magnitude between $10$ and $200$\,mK. Remarkably, at $10$\,mK, $\tau_2$ reaches one month.

%To minimize the impact of probe pulses on the spin density modulation, the probe pulses are sent every half hour (note that even this slow probing rate may have a non-negligible effect on the lowest measured decay rate). We moreover calibrate this impact by recording the echo amplitude as a function of the number of probe pulses in the same conditions of pumping and probing, but without any waiting time between the pulses, and we rescale the measured time dependence by this reference curve (see Fig.~\ref{fig4}) yielding the curve $A_{e,1} (Dt)$. Echoes are still measurable after $5$ days waiting time. The decay is non-exponential, and we phenomenologically fit $A_{e,1} (Dt)$ by a sum of $2$ exponentials with a short (long) time constant $\tau_1$ ($\tau_2$). The constants are shown in Fig.~\ref{fig4}b as a function of the temperature $T$. A strong dependence is observed, with both time constants sharply decreasing when $T$ is increased, changing by two orders of magnitude between $10$ and $200$\,mK. 

We now discuss these results in view of three possible mechanisms that can contribute to the accumulated echo signal decay. The first is the direct relaxation of the nuclear spins involved in the spectral holes, by a $\downarrow_S \uparrow_I - \downarrow_S \downarrow_I$ transition. The rate at which this occurs has no reason to be temperature-dependent in the explored temperature and magnetic field range, and our measurements therefore show that for spins near an Er$^{3+}$ ion, it is at most $3\cdot 10^{-7} \mathrm{s}^{-1}$, more than $6$ orders of magnitude lower than the intra-$^{183}\mathrm{W}$ bath flip-flop rate estimated to be $\approx 1\,\mathrm{s}^{-1}$. Low relaxation rates are indeed expected for nuclei close to a paramagnetic impurity, because the latter produces a magnetic field gradient which detunes these spins from the bath frequency thus slowing down their energy exchange. This so-called frozen core effect~\cite{bloembergen_relaxation_1948} has been observed in numerous systems, including donors in silicon~\cite{wolfowicz_29si_2016,madzik_controllable_2020}, and Nitrogen-Vacancy centers in diamond~\cite{bradley_ten-qubit_2019}; our results show that it is quite pronounced in REI-doped $\mathrm{CaWO}_4$, with a quasi-infinite lifetime for some of the proximal nuclear spins which appear to be completely decoupled from the homonuclear spin bath, likely favored by the low gyromagnetic ratio of $^{183}\mathrm{W}$ nuclei and their low density in scheelite.

A second mechanism is nuclear spin relaxation occurring via the electron spin excited state either resonantly or virtually, analogous to Orbach or Raman electron spin relaxation processes. The Orbach rate should be given by $\sim \Gamma_{1x}/ (\mathrm{e}^{\frac{\hbar \omega_0}{k_B T}} - 1)$, whereas the Raman rate should scale like $T^9$ as in Kramers doublets. These rates moreover depend on the nuclear spin location with respect to the erbium ion, via the cross-relaxation rate. Fitting the Orbach formula to $\tau_1 (T)$ (resp., $\tau_2 (T)$) measurements between $50$\,mK and $220$\,mK with $B$ as only fitting parameter yields reasonable agreement and values of $B/2\pi = 32$\,kHz (resp., $B/2\pi = 124$\,kHz) close to the expected values for nearest-neighbor $^{183}$W nuclei. Therefore, it is likely that the Orbach process is the dominant source of echo decay above $50$\,mK, whereas the Raman process does not seem to match the observations (see Fig.~\ref{fig4}). The two-exponential decay may then originate from the inhomogeneity of the nuclear-spin environment around the Er$^{3+}$ ions, due to the $14\%$ natural abundance of $^{183}$W. The decay times measured at $10$\,mK deviate from the Orbach prediction, indicating either that the sample effective temperature is somewhat higher than the cryostat base temperature (which is quite possible), or that a third mechanism is limiting the hole lifetime. 

%Nearest neighbors (type I and II) $^{183}\mathrm{W}$ have a typical $B/2\pi \approx 50$\,kHz, yielding $\Gamma_{1,x} = 5 \cdot 10^{-3} \mathrm{s}^{-1}$. As seen in Fig.~\ref{fig4} where the predicted $\Gamma_\mathrm{ind}^{-1}(T)$ is shown, the Orbach and Raman processes may contribute in the $100 - 200$\,mK temperature range, but they should become negligible below $100$\,mK where another mechanism seems to take over. 

%Starting from $\downarrow_S \uparrow_I$, a thermal excitation may bring the system into $\uparrow_S \uparrow_I$, followed by cross-relaxation towards $\downarrow_S \downarrow_I$. The rate of this indirect process can be estimated to be $ \Gamma_\mathrm{ind} \approx p_\uparrow (T) \Gamma_{1x} (1+ 2 \bar{n})$, with the excited state probability $p_\uparrow (T) = \mathrm{e}^{-\frac{\hbar \omega_0}{k_B T}}/(1 + \mathrm{e}^{-\frac{\hbar \omega_0}{k_B T}})$, and $\bar{n} = 1/(\mathrm{e}^{\frac{\hbar \omega_0}{k_B T}} - 1)$. $\Gamma_{\mathrm{ind}}$ depends on the temperature as well as on the nuclear spin location with respect to the erbium ion, via the cross-relaxation rate $\Gamma_1 (B / 2\omega_I)^2$. Nearest neighbors (type I) $^{183}\mathrm{W}$ have a typical $B/2\pi \approx 50$\,kHz, yielding $\Gamma_{1,x} = 5 \cdot 10^{-3} \mathrm{s}^{-1}$. As seen in Fig.~\ref{fig4} where the predicted $\Gamma_\mathrm{ind}^{-1}(T)$ is shown, this process may contribute in the $100 - 200$\,mK temperature range, but it appears to be negligible below $100$\,mK where another mechanism seems to take over. 

This third mechanism may be spectral diffusion of the $\mathrm{Er}^{3+}$ spin transition~\cite{mims_spectral_1961}. Whereas the echo amplitude is insensitive to a global $B_0$ drift occurring during the waiting time, a drift of the frequency of each ion, uncorrelated with the others, leads to echo decay. This is in contrast to optical hole burning measurements, which often record the hole area and are therefore insensitive to possible spectral diffusion of the optical line or laser~\cite{konz_temperature_2003}. Spectral diffusion is known to occur in paramagnetic systems, caused for instance by the change in local magnetic field due to spin-flips and flip-flops of surrounding paramagnetic impurities~\cite{klauder_spectral_1962}. Usual methods to measure spectral diffusion involve a stimulated echo sequence~\cite{mims_spectral_1961,rancic_electron-spin_2022,alexander_coherent_2022}, and they probe the drifts on a typical timescale of sub-millisecond to seconds; little is known however on possible spectral diffusion at longer time scales. In that regard, the existence of two time scales might represent several different erbium populations, some with faster spectral diffusion due to their proximity to fluctuating paramagnetic impurities for instance, and some with much slower spectral diffusion. 
%The increased spectral diffusion at higher temperatures may be due to the thermal activation of residual paramagnetic impurities, as already observed in samples with higher $\mathrm{Er}^{3+}$ concentration~\cite{rancic_electron-spin_2022} whose unknown nature, concentration, and resonance frequencies make it unfortunately impossible to evaluate the effect. 
Note that our results imply that the spin frequency of most $\mathrm{Er}^{3+}$ ions drifts by much less than $10$\,kHz ($1$\,ppm relative frequency change) with respect to the center of the line over one month duration at $10$\,mK; to our knowledge, such an exceedingly weak spectral diffusion has never been reported.

Our results show that in ultra-pure scheelite crystals at millikelvin temperatures, nuclear spins around a paramagnetic impurity are completely frozen; in these conditions, the lifetime of spectral holes can be a good probe for spectral diffusion over very long time scales, which is found to be exceedingly weak at the lowest reachable temperature of $10$\,mK. It would be interesting to investigate whether these properties pertain to other crystals and paramagnetic impurities. In principle, similar results could also be obtained in scheelite by replacing the microwave pumping by an optical pumping on the $\mathrm{Er}^{3+}$ $1.5 \mu \mathrm{m}$ transition, which would then open the door to optical storage applications of these ultra-long hole lifetimes. More generally, our measurements suggest that millikelvin temperatures are an interesting and overlooked regime for REI applications, particularly in quantum storage. On that topic, we first note that much greater modulations of the spin density than the one demonstrated here would be required for building an atomic-frequency comb usable for quantum memory, as well as a larger ensemble cooperativity. Assuming these can be achieved, the long hole lifetime would be beneficial for implementing quantum storage protocols in the microwave~\cite{afzelius_proposal_2013,ranjan_multimode_2020} or the optical domain.

%These remarkably long time scales constitute a vivid confirmation of the frozen core concept, and are made possible by the low gyromagnetic ratio of $^{183}\mathrm{W}$ and the low paramagnetic impurities concentration in our $\mathrm{Er}^{3+}:\mathrm{CaWO}_4$ sample. The long hole lifetime and spin coherence also point towards possible applications in atomic-frequency-comb-based microwave quantum memories, in analogy with the optical quantum memories. Such applications would require much stronger spin density modulations than the one achieved in this work, as well as larger spin-ensemble cooperativity coupling to the resonator.

\subsection*{Acknowledgements}
{We acknowledge technical support from P.~S\'enat, D. Duet, P.-F.~Orfila and S.~Delprat, and are grateful for fruitful discussions within the Quantronics group. We acknowledge support from the Agence Nationale de la Recherche (ANR) through the MIRESPIN (ANR-19-CE47-0011) project. We acknowledge support of the R\'egion Ile-de-France through the DIM QUANTIP, from the AIDAS virtual joint laboratory, and from the France 2030 program under the ANR-22-PETQ-0003 grant. This project has received funding from the European Union Horizon 2020 research and innovation program under Marie Sklodowska-Curie grant agreement no. 792727 (SMERC). Z.W. acknowledges financial support from the Sherbrooke Quantum Institute, from the International Doctoral Action of Paris-Saclay IDEX, and from the IRL-Quantum Frontiers Lab. We acknowledge IARPA and Lincoln Labs for providing the Josephson Traveling-Wave Parametric Amplifier. We acknowledge crystal lattice visualization tool VESTA.}

%\subsection*{Author contributions}
%{P.G. grew the crystal, which M.L.D., Z.W. and S.B. characterized through CW and pulse EPR measurements. M.L.D., D.V., P.B., E.F. designed the spin resonator. M.L.D., Z.W. fabricated the spin resonator. M.R. designed and installed the magnetic field stabilization. S.L. and R.B.L. did the simulation. Z.W., M.L.D. took the measurements. Z.W., P.B., analyzed the data. Z.W., P.B., D.V., E.F. wrote the article, with contributions from all the authors. E.F. and P.B. supervised the project.}

\bibliographystyle{naturemag}
\bibliography{main}

\clearpage

\end{document}

%\linenumbers

\title{Method}

\date{\today}

\maketitle

\subsection{Sample}
The $\mathrm{CaWO_4}$ crystal used in this experiment originates from a boule grown by the Czochralski method from $\mathrm{CaCO_3}$ (99.95\% purity) and $\mathrm{WO_3}$ (99.9 \% purity). A sample was cut in a rectangular slab shape ($9~\mathrm{mm}\times4~\mathrm{mm}\times0.5~\mathrm{mm}$), with the surface approximately in the $(ac)$ crystallographic plane, and the $c$-axis parallel to its long edge. The residual doping concentration of erbium is 3.1 $\pm$ 0.2 ppb, measured from continuous-wave EPR spectroscopy \cite{le_dantec_electron_2022}.

\begin{figure}
\centerline{\includegraphics[width=0.5\textwidth]{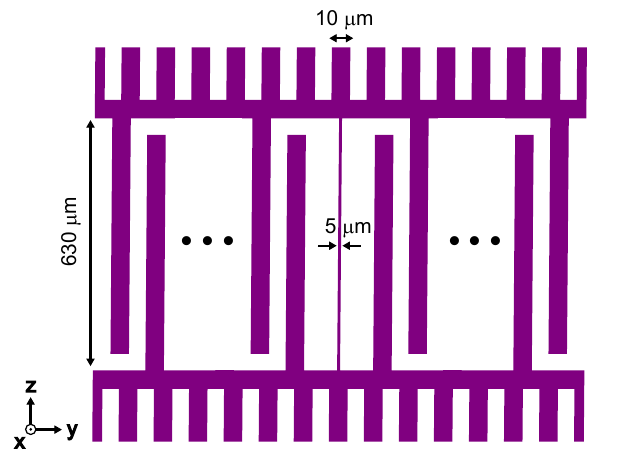}}
\caption{
\textbf{Schematic of the sample design.} A lumped-element LC resonator is formed from a micro-sized inductive wire in the middle and interdigitated finger capacitors on both sides. Each side has 8 pairs of fingers. }
\label{fig_method_sample}
\end{figure}

On top of this substrate, a lumped-element LC resonator, see Extended Data Fig. \ref{fig_method_sample}, was fabricated by sputtering 50 nm of niobium and patterning the film by electron-beam lithography and reactive ion etching. The sample is placed in a 3D copper cavity with a single microwave antenna and SMA port used both for the excitation and the readout in reflection. The top and bottom capacitor pads are shaped as parallel fingers in an effort to improve the resonator resilience to an applied residual magnetic field perpendicular to the metallic film. In the main text, $xyz$ axes are defined as follows (see also Extended Data Fig.\ref{fig_method_sample}). The $z$ axis is defined as being along the wire and $y$ axis being perpendicular to it. The $y$ and $z$ axes are are in the sample plane, while the $x$ axis is perpendicular to the plane. From the approximate sample orientation, the $y$ axis makes a small angle $\beta$ with the crystalline $c$ axis, while the $yz$ plane approximately corresponds to the crystalline $ab$ plane.

\subsection{Experimental setup}
\begin{figure*}
\centerline{\includegraphics[width=0.85\textwidth]{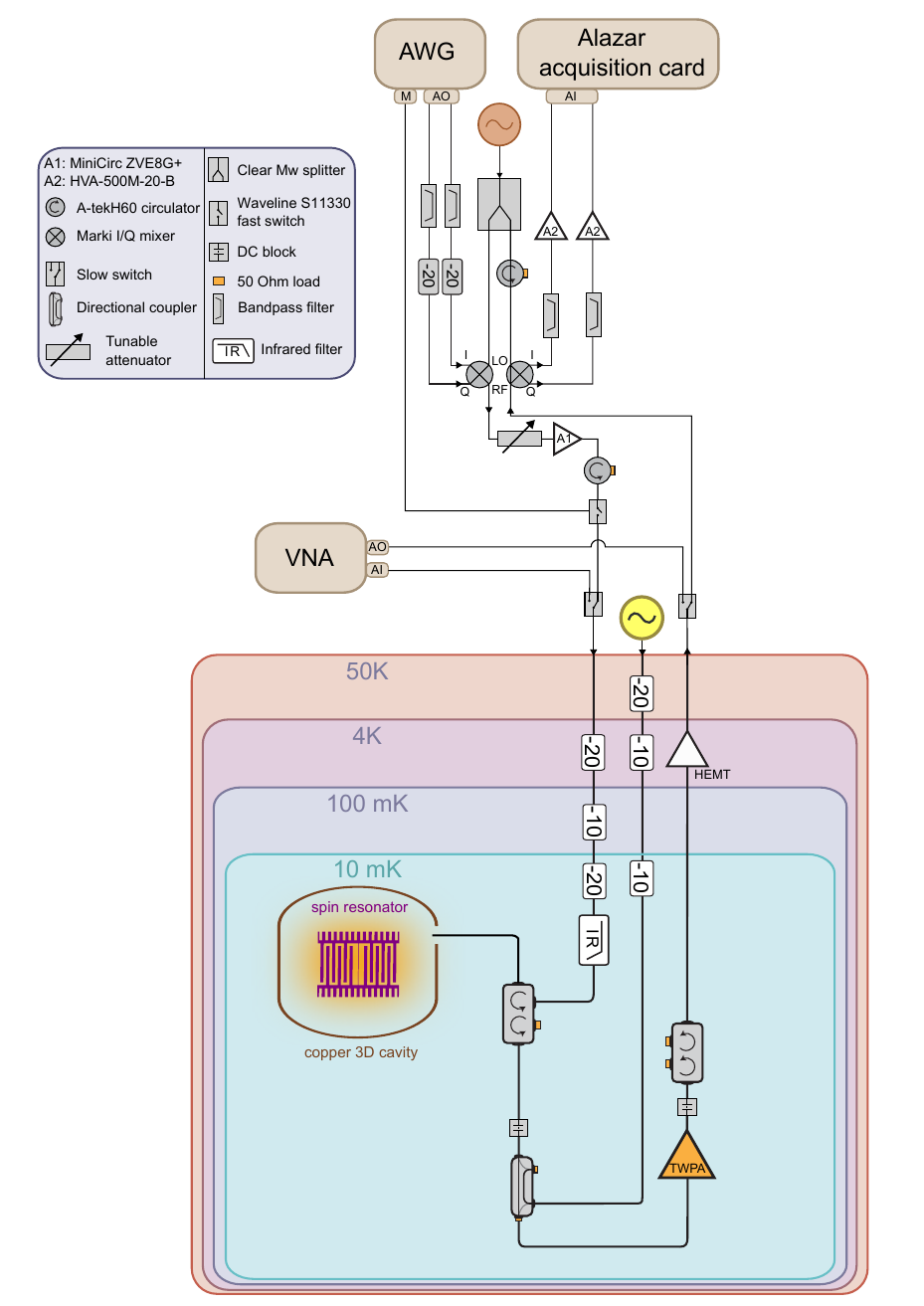}}
\caption{
\textbf{Schematic of the setup.} Wiring and all the components used in this experiment at room temperature and cryogenic temperature are shown. }
\label{fig_method_setup}
\end{figure*}

The complete setup schematic is shown in Extended Data Fig.\ref{fig_method_setup}.

\subsubsection{Room-temperature setup}
The room-temperature setup contains two sets of instruments: (i) a vector network analyser (VNA) for frequency domain measurements, (ii) a microwave source, an arbitrary waveform generator (AWG) and an acquisition card for pulsed time-domain measurements. Two slow switches are used to allow the switching between the sets of VNA spectrum and spin-echo measurements.

The pulses used to drive the spins are generated by mixing a pair of in-phase (I) and quadrature (Q) signals from the AWG with the local oscillator (LO) at the spin resonator frequency $\omega_0$ from a microwave source. 

The demodulation of the output of the signal is achieved by a second I/Q mixer with the same LO. Then the I and Q signals are amplified and digitized by an acquisition card.

\subsubsection{Low-temperature setup}

The spin excitation pulses at the input of the dilution unit are filtered and attenuated to minimize the thermal noise. They are directed,  through a circulator, to the antenna of the 3D cavity containing the spin resonator. The reflected and output signals on this antenna are routed to room temperature through a Josephson Traveling Wave Parametric Amplifier (TWPA), a double-junction circulator for isolation and a HEMT amplifier.

\subsection{Magnetic field alignment and stabilization}
\begin{figure}
\centerline{\includegraphics[width=0.5\textwidth]{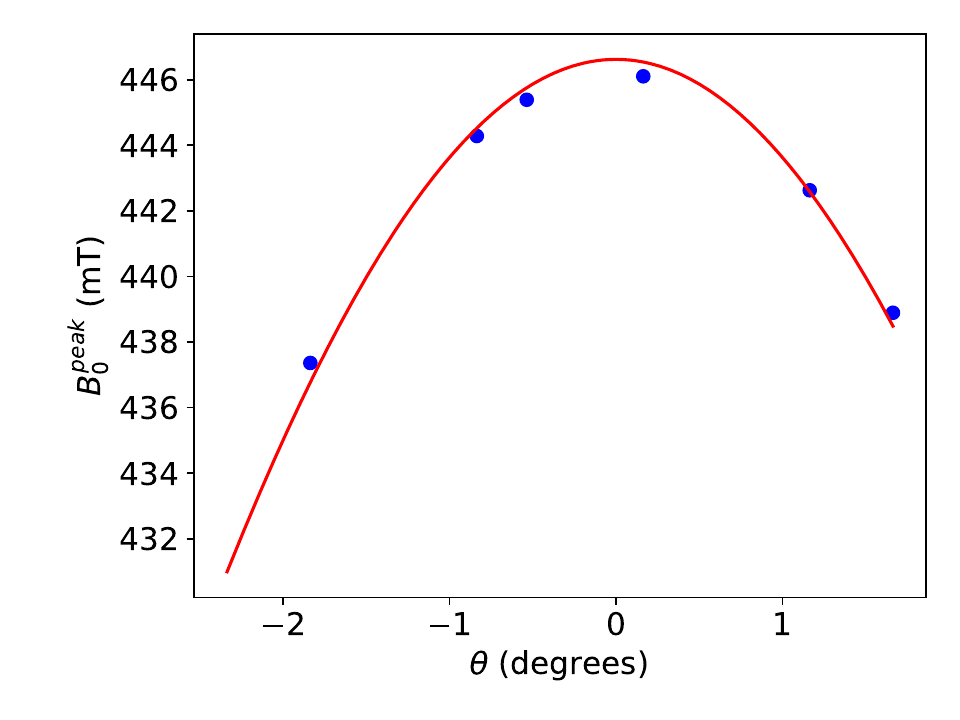}}
\caption{
\textbf{Alignment to c-axis projection in the resonator plane.} Measured (dots) magnetic field  $B_0^{peak}$ at which the center of the spin ensemble line is found, as a function of the angle $\theta$ that the field makes with the $c$ axis projection in the resonator plane. A fit  with eq. \ref{eq:field_alignment} (line) to the data yields the $\theta=0$ origin.}
\label{fig_method_field_alignment1}
\end{figure}

A 1T/1T/1T $3$-axis superconducting vector magnet provides the static magnetic field $B_0$ for this experiment. The alignment of the magnetic field occurs through a two-step process. First, we align the field in the sample plane ($y-z$) by applying a minor field strength of $50$\,mT while minimizing the resonator losses and frequency shift relative to the zero-field values. Second, we ascertain the direction of the crystallographic $c$-axis projection on the sample plane, defined as $\theta=0^{\circ}$, by measuring the erbium spectroscopic ensemble line using a Hahn-echo sequence for various $\theta$ angles within the ($y-z$) plane~\cite{le_dantec_electron_2022}, see Extended Data Fig.\,\ref{fig_method_field_alignment1}. Due to the anisotropy of the gyromagnetic tensor, the center of the ensemble line $B^{peak}_0$ can be expressed with angle $\theta$ and a small field offset $B_\mathrm{offset}$ as

\begin{equation}
    B_0^{peak}=\hbar\omega_0/\sqrt{\gamma_{\parallel}^2 \cos^2 \theta+\gamma_{\perp}^2\sin^2\theta}+B_\mathrm{offset}.
\label{eq:field_alignment}
\end{equation}

At each angle, we scan the total field $B_0$ by simultaneously adjusting the current in all three coils to maintain the same angle. It's worth noting that, due to the absence of a method to determine the slight angle between the $c$-axis projection on the $yz$ plane and the $y$ axis, we used a zero value by simplicity, given that this small residual angle has negligible impact on any of the results found in the article.

Due to imprecision in dicing the $\mathrm{CaWO_4}$ boule, the crystallographic $c$-axis deviates slightly from the sample plane, which is characterized by the angle $\beta$. After the field $B_0$ is well-aligned with $c$-axis projection within the sample plane, a small additional out-of-plane field $\delta B_\perp$ can be introduced to adjust $B_0$ closer to the actual $c$-axis orientation. However, this adjustment comes at the expense of increased internal loss in the resonator, as well as decreased signal-to-noise ratio. The misalignment angle can be obtained from $\beta \sim \mathrm{arctan}(\delta B_\perp/B_0)$ 

\begin{figure}
\centerline{\includegraphics[width=0.5\textwidth]{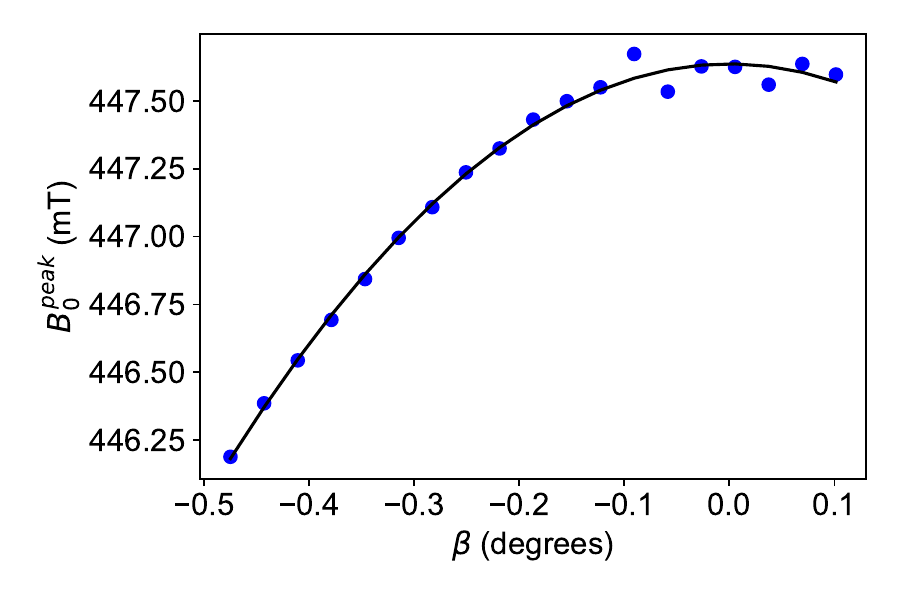}}
\caption{
\textbf{Alignment to c-axis with extra out-of-plane field.} Measured (dots) magnetic field  $B_0^{peak}$ at which the center of the spin ensemble line is found, as a function of the angle $\beta$ that the field makes with the $c$ axis. A parabolic fit (line) to the data yields the $\beta=0$ origin}
\label{fig_method_field_alignment2}
\end{figure}

At each $\beta$ value, we measured the internal loss of the resonator as a function of $B_0$ by probing the reflected signal to obtain the erbium spectroscopic ensemble line, whose center $B_0^{peak}(\beta)$ reaches the maximum value when the total field is aligned to the c-axis, as shown in Extended Data Fig.\,\ref{fig_method_field_alignment2}. This adjustment, however, results in the resonator's internal loss increasing by a factor of 30 compared to the case of $\delta B_\perp=0$. 

In the spectral hole burning measurements ( Fig. 2 in the main text), better alignment with actual $c$-axis provides a more symmetric and straightforward condition for studying the system's hyperfine structure. Therefore, we choose $\delta B_\perp=6$\,mT, as it still allows for a decent signal-to-noise ratio. In the accumulated echo measurements (Fig. 3 and Fig. 4 in the main text), we set $\delta B_\perp=0$ and keep $B_0$ within the sample plane.

The stability of the magnetic field is determined by the operational mode (either current supplied or persistent mode) of the three superconducting coils within the vector magnet, as well as their associated current sources. In the spectroscopy data (Fig. 1) presented in the main text, we utilize the current supplied mode alongside a commercial current source (Four-Quadrant Power Supply Model 4Q06125PS from AMI). Conversely, the SHB and accumulated echo data depicted in Figs. 2, 3 and 4 necessitate a better stability (with less than $10 \mathrm{kHz}$ variation) over long periods of time. Therefore we place all three coils in persistent mode to minimize the noise.

\subsection{Estimation of spin density}
In this section, we derive the expression for estimating the ratio of spin density between the reference and hole burning spectra from the measured reflection coefficient $S_{11}$.

%When the resonator is coupled to the spin ensemble in the crystal, an additional loss is introduced into the system. Assuming the transition frequencies of spin ensemble have a Lorentzian distribution centered at $\omega_s$ with an inhomogeneous linewidth of $\Gamma_{\mathrm{inh}}$, the new total damping rate is given by
%\begin{equation}
%    \widetilde{\kappa_t}=\kappa_c+\kappa_i+\frac{g^2_{\mathrm{inh}}\Gamma_{\mathrm{inh}}}{(\omega-\omega_s)^2+(\Gamma_\mathrm{inh})^2/2},
%\end{equation}
%where $g_{\mathrm{ens}}$ represents the ensemble coupling strength.

The reflection coefficient of the spin-resonator system is 
\begin{equation}
    S_{11}=1-\frac{i \kappa_c}{(\omega-\omega_0)+i (\kappa_c + \kappa_i)/2 - W(\omega)},
\end{equation}

\noindent with $W(\omega) = g_{ens}^2 \int \frac{\rho(\omega') d\omega'}{\omega - \omega' + i \gamma/2}$, $g_{ens} = \sqrt{\int g^2 \rho(g) dg}$ being the ensemble coupling constant, and $\gamma$ the spin homogeneous linewidth. In the limit where $\gamma$ is small compared to all characteristic couplings of the system, it can be shown that $\rho(\omega) = - Im[W]/(\pi g_{ens}^2)$ ~\cite{grezes_multimode_2014}. 

In the limits $\omega\sim\omega_s$ and $\omega\sim\omega_0$,  the reflection coefficient simplifies to:
\begin{equation}
    S_{11}  =  1-\frac{2\kappa_c}{\kappa_i+\kappa_c} + i \frac{4 \kappa_c}{(\kappa_c + \kappa_i)^2} W(\omega).
\end{equation}

We develop $|S_{11}|^2$, keeping only first order terms in $W(\omega)$, and get 

\begin{align}
    |S_{11}|^2 \sim (\frac{\kappa_i-\kappa_c}{\kappa_i+\kappa_c})^2+\frac{4\kappa_c(\kappa_i-\kappa_c) Im[W(\omega)]}{(\kappa_i+\kappa_c)^3}.
\end{align}

Finally, the spin density for probe frequencies closed to the resonance can be obtained from
\begin{equation}
    \rho(\omega)\propto Im[W](\omega) = \frac{(\kappa_i+\kappa_c)^3}{4\kappa_c(\kappa_i-\kappa_c)}\left[|S_{11}(\omega)|^2-(\frac{\kappa_i-\kappa_c}{\kappa_i+\kappa_c})^2\right].
\end{equation}
This leads to the ratio of spin densities between the hole-burning condition ($\rho_1$) and reference condition ($\rho_0$) as follows:
\begin{equation}
    \frac{\rho_1(\omega)}{\rho_0(\omega)}=\frac{|S_{11}^{(1)}(\omega)|^2-(\frac{\kappa_i-\kappa_c}{\kappa_i+\kappa_c})^2}{|S_{11}^{(0)}(\omega)|^2-(\frac{\kappa_i-\kappa_c}{\kappa_i+\kappa_c})^2}.
\end{equation}
Note that $\kappa_i$ and $\kappa_c$ are intrinsic resonator internal loss and external coupling rate when spins are detuned from resonance $\omega_0$.

\subsection{Spin density reset}
Given the long lifetime of the holes imprinted in the spin density, originating from dynamical polarization of nuclear spins in our system, it is crucial to reset the spin density to its equilibrium value. This ensures the removal of the imprinted spectrum pattern from the previous measurement, guaranteeing that each new measurement starts under the same conditions. The reset process can be achieved by sweeping the pump tone with a strong power across a wide frequency range (at least 10 times larger than the scanning range of interest for observing holes). Specifically, we use VNA scan for the reset with a chosen power of -71 dBm at the sample input and a range of 5 MHz centered at $\omega_0$. The process contains 24 VNA scans, each lasting 10 seconds with a step size of 0.5 kHz.

\begin{figure}
\centerline{\includegraphics[width=0.5\textwidth]{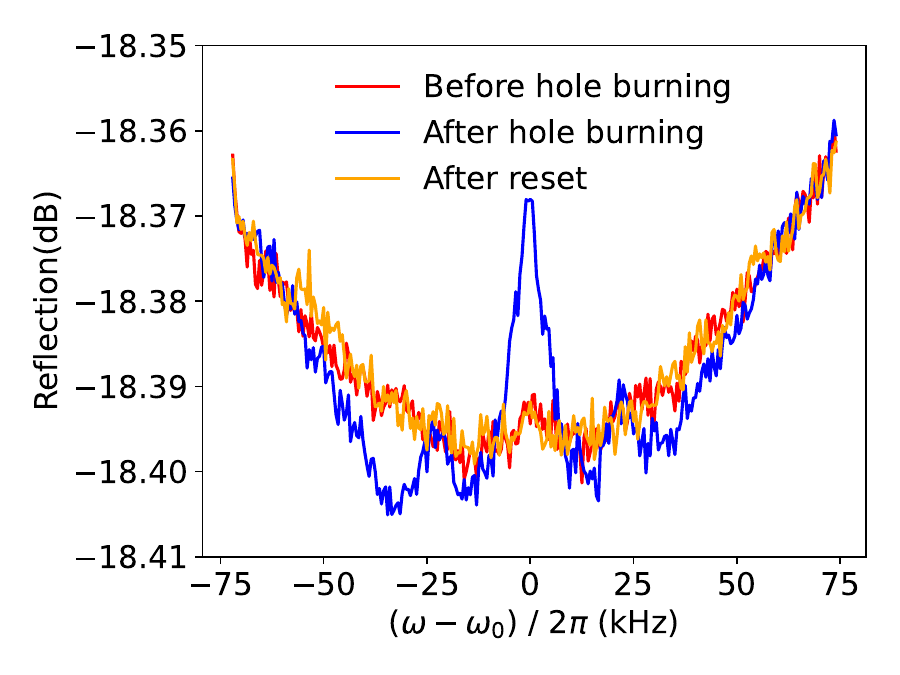}}
\caption{
\textbf{Spin density reset.} Measured reflection spectra of the resonator measured with low power for three cases: before hole burning (red solid line), after a hole has been burned at $\omega_0$ (orange solid line) and after the reset process (orange solid line).}
\label{fig_method_reset}
\end{figure}

To illustrate the efficacy of the reset process, we use low-power VNA scans ($\sim -131$\,dBm at the sample input) to investigate the reflection within the range of interest under various senarios. We first obtain a spectrum to establish a baseline. Subsequently, a hole is created at $\omega_0$ with resonant drive, resulting in a noticeable modification of the spectrum due to the change of spin density. Finally, following the application of the reset process mentioned above, we acquire another spectrum for comparison with the baseline and the spectrum featuring the hole, which demonstrates the erasure of the hole pattern in the spectrum, as shown in Extended Data Fig. \ref{fig_method_reset}.

\subsection{Probe-induced redistribution and signal rescaling}
We now discuss the backaction of the excitation pulses used to retrieve the accumulated echo signal in time domain from the periodic modulation of the spin density. Due to cross-relaxation, the pulse applied to spin ensemble can create depolarization of nuclear spins, which reduces the modulation depth of the grating and therefore affects the measured echo amplitude over time. As a result, the corresponding observed decay time constant becomes shorter than its intrinsic value.

\begin{figure}
\centerline{\includegraphics[width=0.5\textwidth]{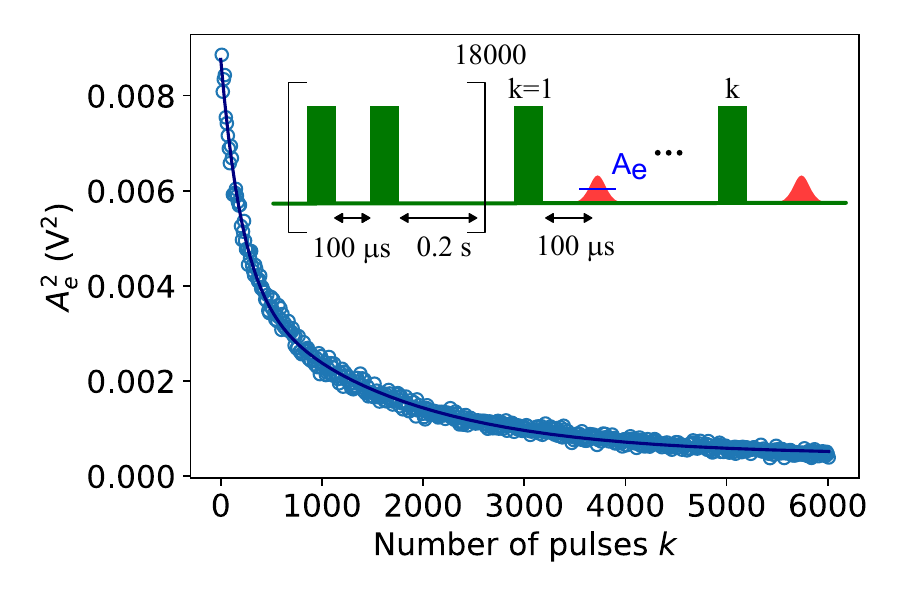}}
\caption{
\textbf{Pulsed induced decay of accumulated echo.} Measured averaged echo energy as a function of number of pulses applied (dot). A triple exponential function is used for fitting (solid line).}
\label{fig_method_rescaling}
\end{figure}

To study the impact of the pulse-induced decay, we measure the echo amplitude as a function of number pulses under identical conditions of pumping ($N=18000$ pairs of pulses separated by $\tau = 100 \mu \mathrm{s}$) and probing, but without any extra waiting time between the pulses. Extended Data Fig.\,\ref{fig_method_rescaling} shows the measured average amplitude of the first echo that appears after the excitation pulse, plotted as a function of the number of pulses applied to the system. This observed pulse-induced decay curve can serve as a reference for calibrating measurements where a waiting time is introduced between probing pulses.

Fitting the measured data to a triple exponential function and rescaling its starting point (no pulse applied) to one provides a scaling factor for the actual measurements with varying number of probing pulses and different waiting times between two adjacent pulses.

\subsection{SHB spectra with misalignment of crystalline c-axis}
\begin{figure}
\centerline{\includegraphics[width=0.5\textwidth]{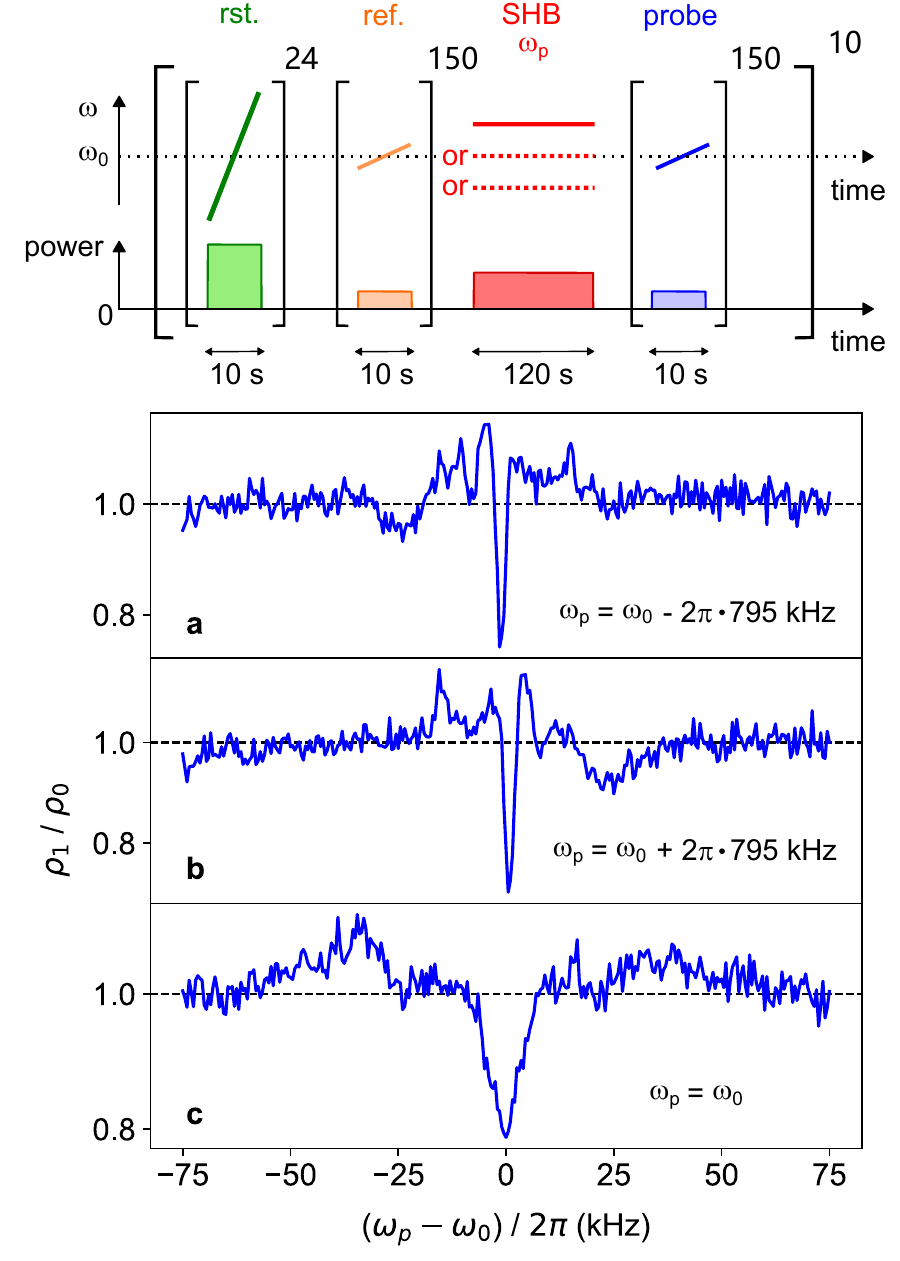}}
\caption{
\textbf{Spectral hole burning with misalignment.} Measured spin density $\rho_1(\omega)$ normalized to the unperturbed spin density $\rho_0(\omega)$, with a misalignment between crystal c-axis and magnetic field $B_0$. Each sequence starts with a repetitive high power ($\sim -71$\,dBm) reset scan with a range of 4 MHz centered at $\omega_0$, followed by a lower power ($\sim -131$ \,dBm) reference scan yielding $\rho_0(\omega)$ , a pumping step at single frequency $\omega_p$ (power$\sim -81$ dBm), and a measurement scan yielding $\rho_1(\omega)$. The entire sequence is repeated for 10 times. The pump frequency $\omega_p/2\pi$ is $\omega_0 /2\pi - 795$\,kHz (a), $\omega_0 /2\pi + 795$\,kHz (b), and $\omega_0$ (c).}
\label{fig_method_hole_burning_misalign}
\end{figure}

In addition to the SHB spectra presented in the main text, where the magnetic field $B_0$ is aligned with the crystal c-axis, we have also recorded the spectra under a slight misalignment, where $B_0$ is in the sample plane but misaligned with the actual c-axis. The same measurement sequence is applied to obtain the 3 spectra for a pump applied at respectively $\omega_0 + \omega_I$  (forbidden transition pumping, on the red sideband), $\omega_0 - \omega_I$ (forbidden transition pumping, on the blue sideband), and $\omega_0$ (allowed transition pumping). The relative spin density change, $\rho_1(\omega)/\rho_0(\omega)$, is plotted in Extended Data Fig.~\ref{fig_method_hole_burning_misalign} for the $3$ pump frequencies. 

\subsection{Decay of hole and spin density modulation spectra}
\begin{figure}
\centerline{\includegraphics[width=0.5\textwidth]{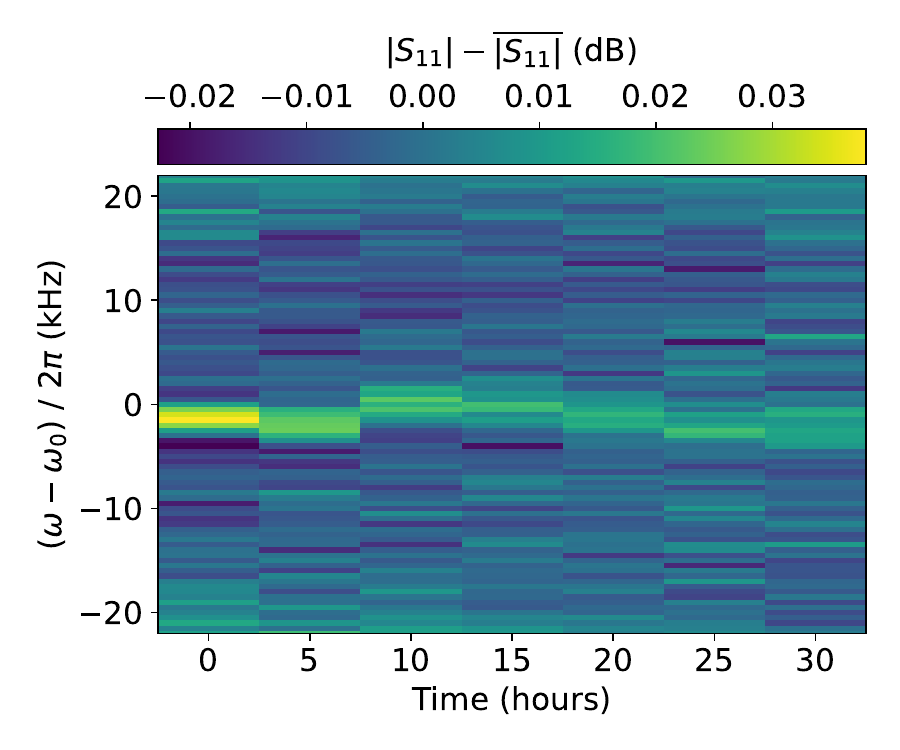}}
\caption{
\textbf{Decay of spectral hole.} Measured reflection coefficient of the resonator (with its mean value subtracted) at different times after a spectral hole has been created.}
\label{fig_method_decay_hole_spectra}
\end{figure}

As a complementary measurement of the decay of the accumulated echo obtained in time domain in the main text, it is also of interest to demonstrate the time evolution of the spectra of both the hole and the periodically modulated pattern. In this measurement, the system is reset, followed by the creation of a hole via the lower forbidden transition using the same protocol as in Extended Data Fig. \ref{fig_method_hole_burning_misalign}. The spectrum is then probed every five hours, as depicted in Extended Data Fig. \ref{fig_method_decay_hole_spectra},  showing the temporal spectral decay featuring a slow fading characteristic. It's noteworthy that, in contrast to rapid pulsed measurements, frequency domain scanning operates at a slower pace with more averaging for noise reduction and thus induces stronger decay due to cross-relaxation.

\begin{figure}
\centerline{\includegraphics[width=0.5\textwidth]{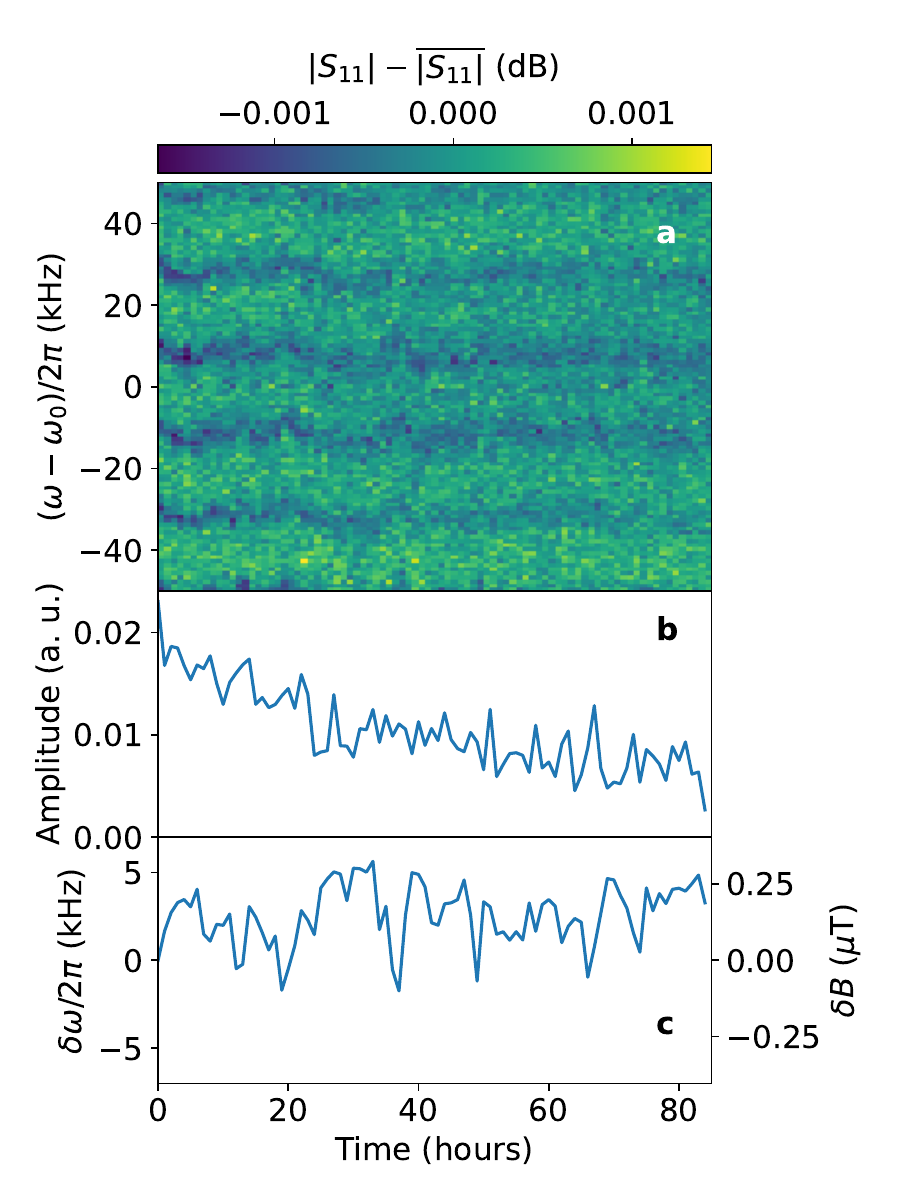}}
\caption{
\textbf{Decay of spin density grating.} \textbf{a}. Measured reflection coefficient of the resonator (with its mean value subtracted) at different times. \textbf{b}. Amplitude of the 20 kHz peak obtained from a fast Fourier transform (FFT) on the dataset in panel \textbf{a}. \textbf{c}. Fluctuation of frequency and corresponding magnetic field obained from the phase of FFT on the dataset in panel \textbf{a}.}
\label{fig_method_decay_AFC_spectra}
\end{figure}

Regarding the spin density modulation, after resetting the system, the grating is created using a similar approach in Extended Data Fig. \ref{fig_method_rescaling}, with a different delay time ($\tau=50\,\mu$s). The probing is performed in the frequency domain with a VNA instead of accumulated echo measurements in the time domain. Extended Data Fig. \ref{fig_method_decay_AFC_spectra}a show the time evolution of the grating pattern up to 80 hours. The separation of two adjacent peaks is 20 kHz as expected from 1/$\tau$ given the chosen $\tau=50 \mu$s. Extended Data Fig. \ref{fig_method_decay_AFC_spectra}b, c present the Fourier analysis for the measured periodic pattern. The amplitude of the 20 kHz peak, obtained from fast Fourier transform (FFT), decays but remains non-zero 80 hours after the creation. The time dependence of FFT phase reveals frequency fluctuations, demonstrating high frequency stability throughout the entire measurement duration. This indicates that the frequency drift due to global magnetic field drift and spectral diffusion is less than 10 kHz, and the corresponding field fluctuation is less than 0.5 $\mu $T.

%\subsection{Effect of modulated spin density on input pulses}

%\begin{figure}
%\centerline{\includegraphics[width=0.5\textwidth]{figures_method/FFT.png}}
%\caption{
%\textbf{Temporal and frequency response of a square pulse from a grating transfer function.}}
%\label{fig_method_pulse_fft}
%\end{figure}

\subsection{Simulation of spectral hole burning and accumulated echo}
\subsubsection{System Hamiltonian}
The system we consider is a single Er$^{3+}$ ion with zero-nuclear-spin
isotope doped in CaWO$_{4}$. The Hamiltonian that describes the subspace of
the ground $^{4}I_{15/2}$ multiplet ($S_{4}$ point symmetry) is given by 
\begin{equation}
H_{Er}=H_{cf}+H_{Z},  \label{H}
\end{equation}%
where $H_{cf}$ is the reduced crystal-field Hamiltonian%
\begin{eqnarray}
H_{cf} &=&\alpha _{J}B_{2}^{0}O_{2}^{0}+\beta _{J}\left(
B_{4}^{0}O_{4}^{0}+B_{4}^{4}O_{4}^{4}+B_{4}^{-4}O_{4}^{-4}\right)   \notag \\
&&+\gamma _{J}\left(
B_{6}^{0}O_{6}^{0}+B_{6}^{4}O_{6}^{4}+B_{6}^{-4}O_{6}^{-4}\right) ,
\label{H_cf}
\end{eqnarray}%
and therein $O_{l}^{m}$ are the Stevens equivalent operators, $\alpha _{J}$, 
$\beta _{J}$, $\gamma _{J}$ are the Stevens coefficients \cite{stevens1952},
and $B_{l}^{m}$ are the crystal-field parameters \cite{enrique1971optical}. $%
H_{Z}=g_{J}\mu _{B}\mathbf{J\cdot B_{0}}$ is the Zeeman interaction, where $%
J=15/2$, $g_{J}=6/5$ is the Land\'{e} factor of Er$^{3+}$, and $\mu _{B}$ is
Bohr magneton. The lowest two energy levels of the Er$^{3+}$ ion under the
magnetic field $\mathbf{B_{0}}$ can be effectively described by means of a
spin-$1/2$ $\mathbf{S}$ Hamiltonian%
\begin{equation}
H_{eff}={\mu _{B}}\mathbf{B_{0}}\cdot \mathbf{g}\cdot \mathbf{S},
\label{H_eff}
\end{equation}%
where the anisotropic g-factor tensor $\mathbf{g}$ has a diagonal form in
the crystal frame, with $g_{\perp }=g_{aa}=g_{bb}=8.38$ and $g_{\parallel
}=g_{cc}=1.247$ \cite{antipin_a._paramagnetic_1968}. While, the existence of
crystal field can offer an additional dipole moment that deviates the
expectation values of the momentum $\mathbf{J}$, which cannot be obtained
from the ones of $\mathbf{S}$ via $H_{eff}$. Therefore, we always adopt the
expectation values of $\mathbf{J}$ instead of $\mathbf{S}$ in later
computations.

The CaWO$_{4}$ bath consists of $^{183}$W nuclear spin bath ($I_{i}=1/2$ and
a total of $N_{s}$ spins) with a natural concentration $p_{n}=0.145$. The
Hamiltonian of the bath is given by 
\begin{equation}
H_{n}={g_{n}\mu _{n}}\sum_{i=1}^{N_{s}}\mathbf{B_{0}}\cdot \mathbf{I}_{i},
\label{H_n}
\end{equation}%
where $g_{n}$ is the g-factor of $^{183}$W nuclear spins and $\mu _{n}$ is
the nuclear magneton. To simplify the computation for hole burning and
accumulated echo, we disregard the internal dipolar interaction among
nuclear spins and focus solely on a nuclear spin bath that does not interact
with each other. The dipolar hyperfine interaction is assumed as%
\begin{equation}
H_{int}=\sum_{i=1}^{N_{s}}\mathbf{J}\cdot {\mathbb{A}}_{i}\cdot \mathbf{I}%
_{i}  \label{H_int}
\end{equation}%
where ${\mathbb{A}}_{i}=\frac{\mu }{4\pi r_{i}^{3} }g_{e}\mu _{B}g_{n}\mu
_{n}\left( 1-3\mathbf{r}_{i}\mathbf{r}_{i}/r_{i}^{2}\right) $, and therein $%
\mu $ is the vacuum permeability, $g_{e}$ is electron g-factor, and ${%
\mathbf{r}}_{i}$ is the displacement between the $i$-th $^{183}$W nuclear
spin and central spin.

We set the $z$-axis as along the direction of $\mathbf{B_{0}}$. Since the
energy gap between the first-excited and ground states of the electron is
much stronger than the interaction from the bath, the Hamiltonian for Er$%
^{3+}$ ion dopped in the CaWO$_{4}$ bath can effectively be described by a
pure dephasing model%
\begin{equation}
H\approx |e\rangle \langle e|\otimes H_{e}+|g\rangle \langle g|\otimes H_{g}
\label{H_eg}
\end{equation}%
with the central-spin-conditional bath Hamiltonian%
\begin{equation}
H_{k}=\omega _{0,k}+H_{n}+\left\langle J_{z}\right\rangle
_{k}\sum_{i=1}^{N_{s}}\left(
A_{i}^{xz}I_{i,x}+A_{i}^{yz}I_{i,y}+A_{i}^{zz}I_{i,z}\right) ,  \label{Hk}
\end{equation}%
where the subscript $k=e\ $($g$) respectively refers to the first-excited
(ground) subspace. $\omega _{0,k}$ and $\left\langle J_{z}\right\rangle _{k}$
are the corresponding electronic level and expectation value of electronic
spin. The energy gap of the two levels, $\omega _{0}=\omega _{0,e}-\omega
_{0,g}$, is referred to as the frequency of resonant transition. $I_{i,x}$, $%
I_{i,y}$, and $I_{i,z}$ are the three components of the vector $\mathbf{z}%
\cdot {\mathbb{A}}_{i}$ with according hyperfine interaction coefficient $%
A_{i}^{xz}$, $A_{i}^{yz}$, and $A_{i}^{zz}$. Due to the $z$-axis symmetry,
the Hamiltonian (\ref{Hk}) can also be transformed into%
\begin{equation}
H_{k}=\omega _{0,k}+\sum_{i=1}^{N_{s}}\left[ \left( \omega _{I}+\left\langle
J_{z}\right\rangle _{k}A_{i}\right) I_{i,z}+\left\langle J_{z}\right\rangle
_{k}B_{i}I_{i,x}\right] ,
\end{equation}%
where $\omega _{I}=g_{n}\mu _{n}\left\vert \mathbf{B}_{0}\right\vert $ is a
gyromagnetic ratio of $^{183}$W nuclear spin. $A_{i}=A_{i}^{zz}$ and $B_{i}=\sqrt{\left(
A_{i}^{xz}\right) ^{2}+\left( A_{i}^{yz}\right) ^{2}}$
respectively refer to isotropic and anisotropic component of the hyperfine
interaction of $i$th nuclear spin. Therein, we can introduce $\eta
_{i,k}=\left\langle J_{z}\right\rangle _{k}B_{i}/\left( \gamma
_{W}+\left\langle J_{z}\right\rangle _{k}A_{i}\right) $ as a mixing angle
that quantify these two isotropic and anisotropic hyperfine interactions.

With the help of exact diagonalization (ED), the energy spectrum of a
nuclear spin bath with $N_{s}$ spins can be formally written as 
\begin{equation}
H_{k}=\sum_{j=1}^{2^{N_{s}}}\varepsilon _{k_{j}}\left\vert
k_{j}\right\rangle \left\langle k_{j}\right\vert ,
\end{equation}%
where $\varepsilon _{k_{j}}$ ($\left\vert k_{j}\right\rangle $) is the
system eigenenergies (eigenstates) with $j=1,...,2^{N_{s}}$. Each system
eigenstate $\left\vert k_{j}\right\rangle $ is a series multiplication of
each single nuclear spin $\left\vert s_{l}^{k}\right\rangle $ ($l=1,...,N_{s}
$) coupled with electronic state $\left\vert \Uparrow \right\rangle $ ($%
\left\vert \Downarrow \right\rangle $): when $k=e$ ($g$), it gives $%
\left\vert e_{j}\right\rangle =\left\vert \Uparrow \right\rangle \otimes
\left\vert s_{1}^{e}\right\rangle \otimes ...\otimes \left\vert
s_{N_{s}}^{e}\right\rangle $ ($\left\vert g_{j}\right\rangle =\left\vert
\Downarrow \right\rangle \otimes \left\vert s_{1}^{g}\right\rangle \otimes
...\otimes \left\vert s_{N_{s}}^{g}\right\rangle $) and therein $%
s_{l}^{k}=u_{l}^{k}$ or $d_{l}^{k}$ stands for the nuclear eigenstate of up $%
\omega _{k_{l}}/2$ or down $-\omega _{k_{l}}/2$ eigenenergy. The relation
between each nuclear-spin eigenstate has the form $\left\vert \left\langle
u_{i}^{e}\right. \left\vert u_{i}^{g}\right\rangle \right\vert
^{2}=\left\vert \left\langle d_{i}^{e}\right. \left\vert
d_{i}^{g}\right\rangle \right\vert ^{2}=\lambda _{i}\approx 1$ and $%
\left\vert \left\langle u_{i}^{e}\right. \left\vert d_{i}^{g}\right\rangle
\right\vert ^{2}=\left\vert \left\langle d_{i}^{e}\right. \left\vert
u_{i}^{g}\right\rangle \right\vert ^{2}=\xi_{i}\ll \lambda _{i}$. Therefore,
if we consider the transition matrix component $\left\langle
e_{j}\right\vert J_{x}\left\vert g_{l}\right\rangle $, the process of
multi-nuclear-spin-flipping can almost be neglected, since the transition
probability for $n$-spin-flipping process is effectively proportional to $%
\xi ^{n-1}$. For later hole-burning and accumulated echo computations, we
only take into consideration processes involving at most one nuclear-spin
flip (spin-conversed) .

\subsubsection{Generation of hole and anti-hole}

We investigate the spectrum hole burning by means of continuous-wave pumping
to alter the probabilities distribution of energy spectrum and thus generate
the hole and anti-hole burning in transition spectrum. The corresponding
pumping process is described by the driven Hamiltonian%
\begin{equation}
H_{drive}=\sum_{i,j}\omega _{p}\left\langle e_{i}\right\vert J_{x}\left\vert
g_{j}\right\rangle e^{-i\omega _{d}t}\left\vert e_{i}\right\rangle
\left\langle g_{j}\right\vert +H.c. ,  \label{H_drive}
\end{equation}%
where $\omega _{p}$ and $\omega _{d}$ are respectively the pumping amplitude
and frequency. H.c. stands for the hermitian conjugate of the Hamiltonian.

If we focus on the transition between $\left\vert e_{i}\right\rangle $ and $%
\left\vert g_{j}\right\rangle $, the pumping rate equations can be written as%
\begin{equation}
\left\{ 
\begin{array}{c}
{\dot{\rho}_{e_{i}}=-\tilde{\Omega}_{i}\rho _{e_{i}}+\sum_{j}\Omega
_{ij}\rho _{g_{j}},} \\ 
{\dot{\rho}_{g_{j}}=-\tilde{\Omega}_{j}\rho _{g_{j}}+\sum_{i}\Omega
_{ij}\rho _{e_{i}},}%
\end{array}%
\right. \left( i,j=1,2,...,2^{N_{s}}\right) ,  \label{rio1}
\end{equation}%
where $\rho _{e_{i}}=\left\vert \left\langle e_{i}\right. \left\vert
e_{i}\right\rangle \right\vert ^{2}$ and $\rho _{g_{j}}=\left\vert
\left\langle g_{j}\right. \left\vert g_{j}\right\rangle \right\vert ^{2}$ is
the probabilities for the eigenstates $\left\vert e_{i}\right\rangle $ and $%
\left\vert g_{j}\right\rangle $. $\tilde{\Omega}_{i\left( j\right)
}=\sum_{j\left( i\right) }\Omega _{ij}$, and the pumping rate, which based
on the Fermi golden rule, is given by 
\begin{equation}
\Omega _{ij}=2\omega _{p}^{2}\left\vert \left\langle e_{i}\right\vert
J_{x}\left\vert g_{j}\right\rangle \right\vert ^{2}\frac{\Gamma _{2}}{\left(
\Delta _{ij}^{2}+\Gamma _{2}^{2}\right) }.
\end{equation}%
There is a broadening with a Lorentz distribution, and the half-width at
half-maximum (HWHM) $\Gamma _{2}$ stems from $T_{2}=30$ ms. $\Delta
_{ij}=\varepsilon _{e_{i}}-\varepsilon _{g_{j}}-\omega _{d}+\delta $, and
therein $\delta $ is the random central spin detuning and admits the
Gaussian distribution $\delta \sim \frac{\exp \left( -\frac{\delta ^{2}}{%
2\sigma ^{2}}\right) }{\sigma \sqrt{2\pi }}$ with spin broadening linewidth $%
\sigma /2\pi =8$ MHz. This linewidth is much larger than the linewidth of
the resonant cavity (around $1$ MHz). Thus, we assume a uniform detuning in
our simulation.

Moreover, if we also consider the electronic relaxation process, which
induces the transition from $\left\vert e_{i}\right\rangle $ to $\left\vert
g_{j}\right\rangle $, the Eq. (\ref{rio1}) is revised to%
\begin{equation}
\left\{ 
\begin{array}{c}
\dot{\rho}_{e_{i}}=-\tilde{\Lambda}_{i}\rho _{e_{i}}+\sum_{j}\Omega
_{ij}\rho _{g_{j}}, \\ 
\dot{\rho}_{g_{j}}=-\tilde{\Omega}_{j}\rho _{g_{j}}+\sum_{i}\Lambda
_{ij}\rho _{e_{i}},%
\end{array}%
\right. \left( i,j=1,2,...,2^{N_{s}}\right) ,  \label{rio2}
\end{equation}%
where $\tilde{\Lambda}_{i}=\sum_{j}\Lambda _{ij}$ with $\Lambda _{ij}=\Omega
_{ij}+\Gamma _{ij}$. Therein, $\Gamma _{ij}=\Gamma _{1}\left\vert
\left\langle e_{i}\right\vert J_{x}\left\vert g_{j}\right\rangle \right\vert
^{2}$ is the rate of relaxation process and $\Gamma _{1}=1/T_{1} =5$ s$^{-1}$.

Our focus is on the evolution of probability distribution for each level
under the continuous pumping process. Specifically, we extract the diagonal
component of density matrix and redefine a new $2^{N_{s}+1}$dimensional
vector $\left\vert \rho \right\rangle =\left[ \rho _{e_{1}},...,\rho
_{e_{2^{N_{s}}}},\rho _{g_{1}},...,\rho _{g_{2^{N_{s}}}}\right] ^{T}$, which
is spanned in Cartesius basis that denoted as $\left\{ \left\vert
e_{1}\right\rangle ,...,\left\vert e_{2^{N_{s}}}\right\rangle ,\left\vert
g_{1}\right\rangle ,...,\left\vert g_{2^{N_{s}}}\right\rangle \right\} $.
The rate equation of Eq. (\ref{rio2}) can be reformulated as 
\begin{equation}
\frac{\mathrm{d}\left\vert \rho \right\rangle }{\mathrm{d}t}=M\left\vert
\rho \right\rangle ,
\end{equation}%
where $M$ is a $2^{N_{s}+1}$ dimension transition matrix with its non-zero
components giving by%
\begin{equation}
\left\{ 
\begin{array}{c}
\left\langle e_{i}\right\vert M\left\vert e_{i}\right\rangle =-\tilde{\Lambda%
}_{i},\left\langle e_{i}\right\vert M\left\vert g_{j}\right\rangle =\Omega
_{ij}, \\ 
\left\langle g_{j}\right\vert M\left\vert e_{i}\right\rangle =\Lambda
_{ij},\left\langle g_{j}\right\vert M\left\vert g_{j}\right\rangle =-\tilde{%
\Omega}_{j},%
\end{array}%
\right. \left( i,j=1,2,...,2^{N_{s}}\right) .
\end{equation}%
At time $t$, the probability distribution $\left\vert \rho \right\rangle $
can be obtained via the following evolution 
\begin{equation}
\left\vert \rho \right\rangle =e^{Mt}\left\vert \rho _{0}\right\rangle ,
\end{equation}%
where the initial state $\left\vert \rho _{0}\right\rangle $ is usually set
as the thermal state $\left\vert \rho _{0}\right\rangle =\left[
0,...,0,1/2^{N_{s}},...,1/2^{N_{s}}\right] ^{T}$. The changes from $%
\left\vert \rho _{0}\right\rangle $ to $\left\vert \rho \right\rangle $ form
the basis for generating a hole and anti-hole. After the pumping process
around $120$ s, the $T_{1}$ relaxation process will still alter the
probability distribution (denoted by $\left\vert \rho ^{^{\prime
}}\right\rangle $), the components of which are given by 
\begin{equation}
\left\langle e_{j}\right. \left\vert \rho ^{^{\prime }}\right\rangle
=0,\left\langle g_{j}\right. \left\vert \rho ^{^{\prime }}\right\rangle
=\left\langle g_{j}\right. \left\vert \rho \right\rangle +\left\langle
e_{j}\right. \left\vert \rho \right\rangle
\end{equation}

Accordingly, we present the results of hole burning via the relative spin
density change $\rho_{1}(\omega)/\rho_{0}(\omega)$ in Fig. 2 of the main
text, where the spin density is defined as%
\begin{equation}
\rho_{1}\left( \omega \right) =\int \mathrm{d}\delta \sum_{j=1}^{2^{N_{s}}}%
\frac{\gamma \left\vert \left\langle e_{j}\right\vert J_{x}\left\vert
g_{j}\right\rangle \right\vert ^{2}\left\langle g_{j}\right. \left\vert \rho
^{^{\prime }}\right\rangle }{\pi \left[ \left( \varepsilon
_{e_{j}}-\varepsilon _{g_{j}}+\delta -\omega \right) ^{2}+\gamma ^{2}\right] 
},  \label{S_probe}
\end{equation}%
where only the no-spin-flipping transition is considered since it
contributes the most of the spectrum in the probe process. $\gamma$ is the corresponding HWHM, which characterizes each small broadening contributing to the spin density. Meanwhile, we can
also define the reference spin density $\rho_{0}\left( \omega \right) $ by
substituting $\left\vert \rho ^{^{\prime }}\right\rangle $ from Eq. (\ref%
{S_probe}) into $\left\vert \rho _{0}\right\rangle $ in order to observe how
much probabilities have been altered.

The simulation results, which are compared with experimental ones, are shown
in Fig. in the main text. In the experiments, three driven frequencies $%
\omega _{d}=\omega _{L}$, $\omega _{R}$, and $\omega _{0}$ were used to
respectively induce the lower forbidden ($\omega _{L}=\omega _{0}-\omega _{I}
$), higher forbidden ($\omega _{L}=\omega _{0}-\omega _{I}$), and resonant
transitions $\omega _{0}/2\pi =7.808$ GHz, where $\omega _{I}$ is $^{183}$W
nuclear spin Zeeman frequency. The experiments were conducted with the
external magnetic field primarily applied along $z$-axis ($c$-axis), with a
slight bias of about $1$ mT in $y$-axis ($b$-axis), i.e., $\mathbf{B_{0}}%
=(0,1,450)$ mT, which leads to $\left\langle J_{z}\right\rangle _{e}=-0.7$
and $\left\langle J_{z}\right\rangle _{g}=0.3$.

The simulation results are based on an ensemble average of $500$ different
nuclear spin spatial configurations ($N_{s}=8$), which provide results that
are almost in agreement with the experimental data. In the following, we
will discuss more detail about how we obtained the simulation results.

The transition spectrum for $\omega _{L}$ and $\omega _{R}$ driven
situations shows symmetric results about the central frequency $\omega _{0}$%
. For each $N_{s}$-spin bath, the lower (higher) forbidden transition
pumping can theoretically generate $N_{s}$ holes located at $\omega
_{l,hole}=\omega _{L}+\omega _{e_{l}}$ ($\omega _{H}-\omega _{e_{l}}$) and $%
N_{s}$ anti-holes located at $\omega _{l,anti-hole}=\omega _{L}+\omega
_{g_{l}}$ ($\omega _{H}-\omega _{e_{l}}$) in the transition spectrum, where $%
l=1,...,N_{s}$. The experimental result (red curve) in Fig. 2a (b) in the
main text shows an obvious $40$ kHz-interval hole and anti-hole pair, the
central frequency of which is about $-(+)15$ kHz deviated from $\omega _{0}$%
. The hyperfine strength of nuclear spins on the first shell (nearest four)
is much larger than that of other spins, making it the most likely that the
first-shell-spin-flipping process is responsible for the generation of this
hole and anti-hole pair. To obtain agreement with the experimental data, we
need to tune the hyperfine strength of Type I spins, instead of using the dipolar
approximation for hyperfine interaction, which is not applicable when
considering nuclear spins located near the central spin (see Table I). However, the main
hole and anti-holes near the central frequency $\omega _{0}$ are the
compound-spin results of the relatively distant nuclear-spin dynamics.

\begin{table}[th]
\centering%
\begin{tabular}{|c|c|c|c|c|c|}
\hline
$\text{Type}$ & $r_{i} \text{(nm)} $ & $\left\vert A_{i}\right\vert
/2\pi $ & $\left\vert B_{i}\right\vert /2\pi $ & $%
\left\vert A_{i}^{\mathrm{di}}\right\vert /2\pi $ & $\left\vert
B_{i}^{\mathrm{di}}\right\vert /2\pi $ \\\hline
\textrm{I} & $0.3707$ & $73.0/23.0$ & $0$ & $38.7/38.7$ & $0$ \\\hline
\textrm{II} & $0.3867$ & $14.8$ & $35.7$ & $21.2$ & $51.0$ \\\hline
\textrm{III} & $0.5687$ & $21.5$ & $0$ & $21.5$ & $0$ \\\hline
\end{tabular}
\caption{The isotropic and anisotropic hyperfine coefficients of all 10 nearest-neighbor W atoms, classified into three types and shown in Fig. 1a of the main text. $r_{i}$ is the distance from each atom to the Er$^{3+}$ ion, $A_{i}$ ($A_{i}^{\mathrm{di}}$) and $B_{i}$ ($B_{i}^{\mathrm{di}}$) are respectively the fitted (pure dipolar interaction) values, with only those of Types I and II needed to fit. All the hyperfine values are displayed in units of kHz. As for Type I spins, the fitted values depend on the electronic spin state, thus giving two values for each coefficients.
}
\label{table_comparison}
\end{table}
As for the $\omega _{0}$ driven situation, there is one hole at the central
frequency $\omega _{l,hole}=\omega _{0}$ and $2N_{s}$ symmetric anti-holes
at $\omega _{l,anti-hole}=\omega _{0}\pm \omega _{g_{l}}\mp \omega _{e_{l}}$%
. Among them, simulation shows that the nuclear spins on the second shell
(next-nearest) contribute the most to the spectrum. Here we also tune the
hyperfine strength of Type II spins in order to
obtain the results that agree with the experimental data.

Nevertheless, there may be several reasons for the slight inconsistency
between the experimental and simulation results. First, the oversimplified
model used: although the simulation we have considered above can provide the
position of the holes and anti-holes, it cannot precisely predict their
depth or height since it only considers the ensemble average of
eight-nuclear-spin bath. Second, unknown free parameters, such as pumping
amplitude $\omega _{p}$ or the hyperfine strength of nuclear spins on the
first and second shells, were used to tune the simulation. These parameters
may have different values in the experimental setup, leading to
discrepancies between the simulation and experimental results. Further
studies may require a more accurate model and a better understanding of the
experimental parameters.

\subsubsection{Model of polarization transfer and generation of accumulated echo}

In the preceding section, we discussed the probability distribution of the
system energy spectrum, which is modulated by continuous-wave pumping and
results in the hole and anti-hole burning in the transition spectrum.
However, instead of relying solely on continuous-wave pumping, we also conducted experiments using a series of pulse sequences as a polarization generator. This approach creates a modulated spin density and a grating in the transition spectrum. In this section, we demonstrate the generation of the grating through simulations. 
%However, instead of using continuous-wave pumping, additional experiments are also conducted by means of series of pulse sequences which consist of polarization generator, and therefore generates gratings in the transition spectrum. In this section, we provide a demonstration about the generation of the grating from a simulation perspective.

Similar to the hole burning scheme, the system is initially located at the
electronic ground state $\hat{\rho}_{0}=1/2^{N_{s}}\sum_{j=1}^{2^{N_{s}}}%
\left\vert g_{j}\right\rangle \left\langle g_{j}\right\vert $. The
polarization generator used in this experiment, denoted as $\pi /2$-$\tau $-$%
\pi /2$, consists of a two non-selective microwave (m.w.) pulses separated
by a time interval $\tau $. In the rotating frame, which rotates in the
right-hand sense with frequency $\omega _{0}$ about the $z$-axis of the
laboratory frame, the $\pi /2$ pulse is assumed to cause the electron spin
to flip with the angle $\pi /2$ along the direction of m.w. field ($x$%
-axis), with frequency $\omega_{1}$ and duration $t_{p}$, such that $\omega
_{1}t_{p}=\pi /2$. In our simulation, we set $\omega _{1}/2\pi =0.25$ MHz
and $t_{p}=1$ $\mu \mathrm{s}$, which effectively induces only the
nuclear-spin resonant transition but not the lower (higher) forbidden
transition, due to the fact that $\omega _{I}>\omega _{1}$.

For each resonant transition $\left\vert e_{j}\right\rangle \leftrightarrow
\left\vert g_{j}\right\rangle $, the presence of a large electronic-spin
broadening $\delta $ can cause the precession axis to tilt from $\phi =0$ ($x
$-axis) into $\phi _{j}=\arctan \left( \left( \delta +\varepsilon
_{e_{j}}-\varepsilon _{g_{j}}\right) /\omega _{1}\right) $ in the $x$-$z$
plane, with a frequency $\omega _{eff,j}=\sqrt{\left( \delta +\varepsilon
_{e_{j}}-\varepsilon _{g_{j}}\right) ^{2}+\omega _{1}^{2}}$. The matrix form
of $\pi /2$ pulse in the subspace $\left\{ \left\vert e_{j}\right\rangle
,\left\vert g_{j}\right\rangle \right\} $ is defined as $R_{j}=\cos \theta
_{j}\mathbb{I}-i\sin \theta _{j}\left( \sigma _{z}\sin \phi _{j}+\sigma
_{x}\cos \phi _{j}\right) $, where $\theta _{j}=\omega _{eff,j}t_{p}/2$, $%
\mathbb{I}$ is the identity matrix, and $\sigma _{i}$ ($i=x,y,z$) are Pauli
matrix components. Therefore, $\pi /2$ pulse for the whole system is
expressed as $R^{\pi /2}=R_{1}\otimes ...\otimes R_{2^{N_{s}}}$.

\begin{figure}[tbp]
\centerline{\includegraphics[width=0.5\textwidth]{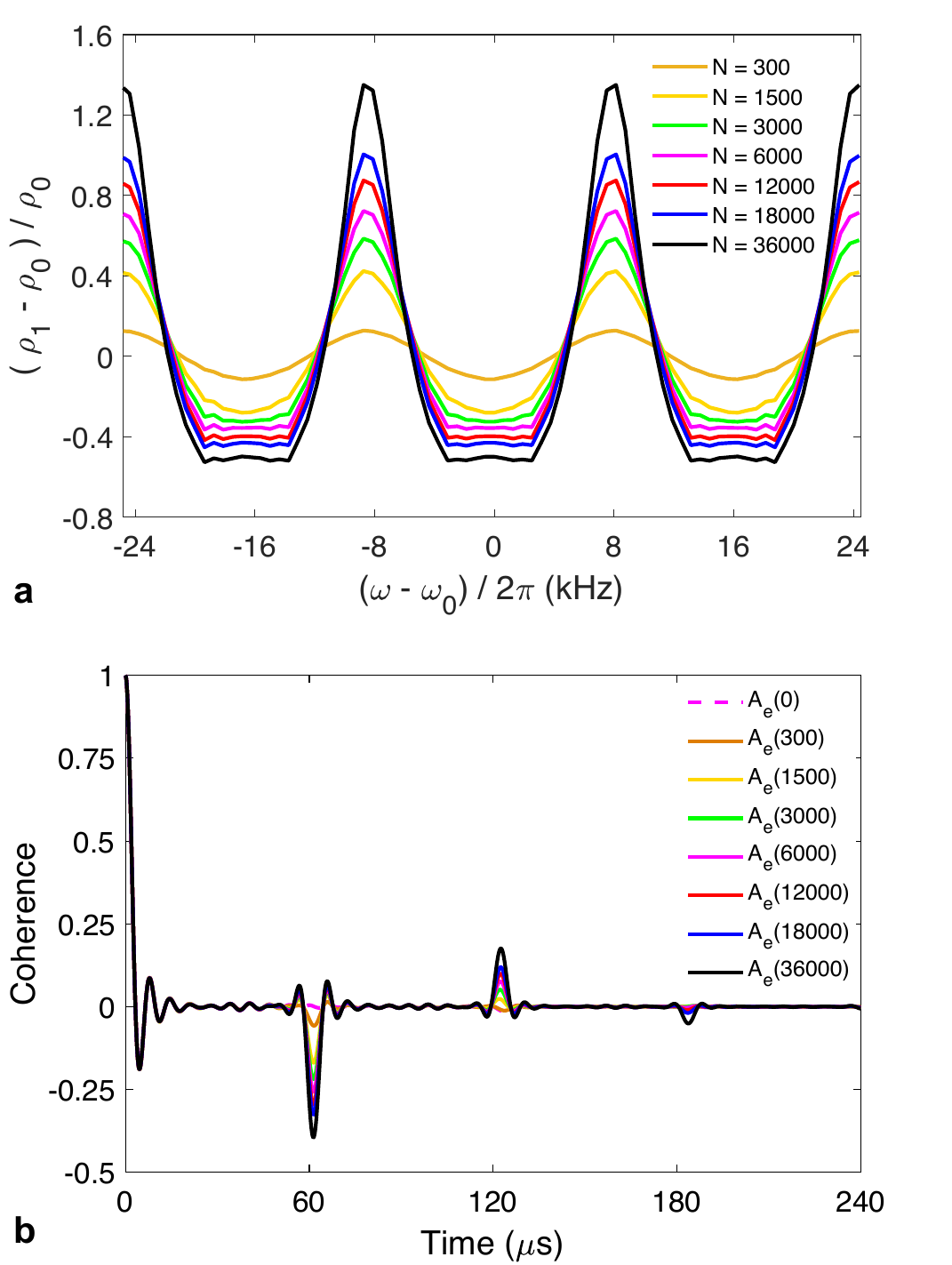}}
\caption{\textbf{The generation and detection of accumulated echo with $\tau =60\,\mu\mathrm{s}$}. \textbf{a}, Relative changes for
the transition spectrum as a function of probe frequency (detuned from $\protect%
\omega _{\text{0}}$) are shown as the formation of the grating, when the system is
subjected to the different sequence numbers of polarization generators with $N=300$, $1500$, $3000$, $6000$, $12000$, $18000$, and $36000$. The
interval between the  nearest two anti-holes is $\Delta \protect\omega /2%
\protect\pi =16.7$ kHz when $\protect%
\tau =60$ $\mathrm{\protect\mu }\mathrm{s}$. \textbf{b}, The coherence obtained after the probe process is used to detect the system probability distribution. As each polarization generator sequence is applied, the echo signals become weaker at $t\approx n\protect\tau $ ($n$ is the positive integer) as time goes on, while the amplitude for each echo signal amplifies with increasing the sequence number $N$. No signal is observed
when $N=0$, which serves as a reference for the
coherence of the initial thermal equilibrium distribution (pink dashed
curve).}
\label{fig_method_AFC_simu}
\end{figure}

In the experiments, the system undergoes a series of polarization generators
with a total sequence number $N$. Before the $n$th sequence, the initial
density matrix is denoted by $\hat{\rho}_{0,n}$, with $\hat{\rho}_{0,1}=\hat{%
\rho}_{0}$ when $n=1$. After each $\pi /2$-$\tau $-$\pi /2$ polarization
generator, the density matrix is modified into 
\begin{equation}
\hat{\rho}_{1,n}=R^{\pi /2}U_{\tau }R^{\pi /2}\hat{\rho}_{0,n}R^{-\pi
/2}U_{\tau }^{\dagger }R^{-\pi /2},  \label{rio_1}
\end{equation}%
where $U_{\tau }$ is the free-evolution operator of the Hamiltonian. There
is a time interval between each two polarization generator sequences, during
which the relaxation and decoherence processes cause the density matrix
becoming $\hat{\rho}_{2,n}=\sum_{j=1}^{2^{N_{s}}}p_{n,j}\left\vert
g_{j}\right\rangle \left\langle g_{j}\right\vert $ with 
\begin{equation}
p_{n,j}=\left\langle g_{j}\right\vert \hat{\rho}_{1,n}\left\vert
g_{j}\right\rangle +\sum_{i=1}^{2^{N_{s}}}\left\vert \left\langle
g_{j}\right\vert \sigma _{x}\left\vert e_{i}\right\rangle \right\vert
^{2}\left\langle e_{i}\right\vert \hat{\rho}_{1,n}\left\vert
e_{i}\right\rangle .  \label{rio_2}
\end{equation}%
After each sequence of polarization generators, the density matrix of the
system changes, with $\hat{\rho}_{0,n+1}=\hat{\rho}_{2,n}$. The final
probability distribution is represented by $\hat{\rho}_{2,N}$ after $N$
sequences of polarization generators, which can form periodical modulated
holes and anti-holes in the transition spectrum, as shown in
Extended Data Fig. \ref{fig_method_AFC_simu}a. By increasing the sequence number $N$, deeper holes
and higher anti-holes can be obtained in the spectrum. The nearest interval
between the two anti-holes is related to the time interval $\tau $ in the
polarization generator, specifically $\Delta \omega =2\pi /\tau $.

Once a grating has been created in the transition spectrum, we can use a $\pi /2$-$\tau $ sequence to detect the accumulated echo or the distribution of $\hat{\rho}_{2,N}$ with the help of
transforming the polarization into coherence. After the probe process,
the density matrix of the system is defined as%
\begin{equation}
\hat{\rho}_{\mathrm{e}}=U_{\tau }R_{x}^{\pi /2}\hat{\rho}_{2,N}R_{x}^{\pi /2\dagger }U_{\tau }^{\dagger },  \label{rio_probe}
\end{equation}%
where the off-diagonal term corresponds to the coherence of the system

\begin{equation}
A_{\mathrm{e}}\left(t, N\right) =\sum_{j=1}^{2^{N_{s}}}p_{N,j}e^{it\left( \varepsilon _{e_{j}}-\varepsilon _{g_{j}}\right) }.
\label{L_probe}
\end{equation}%
The simulation results are displayed in Extended Data Fig. \ref{fig_method_AFC_simu}b, where the echo signals are visible when $t\approx n\tau $ ($n$ is the
positive integer). The amplitudes for each echo decrease over time. As more sequences of pulse are performed on the system, the
corresponding echo signals become amplified. In addition, Fig. 3f in the main text is
referred to as the comparison between the experimental and simulation
results for each sequence $N$, which are in almost agreement.

\bibliography{method}

\clearpage